\documentclass[a4paper,12pt]{article}
\usepackage{amsmath}
\usepackage{amssymb}
\usepackage{amsthm}
\usepackage{latexsym}
\usepackage{mathrsfs}
\usepackage[dvips]{graphicx}
\usepackage[colorlinks=true, citecolor=blue]{hyperref}
\usepackage{pstricks}
\usepackage{color}
\usepackage{tikz}
\usetikzlibrary{decorations.pathreplacing}
\usetikzlibrary{calc}
\usetikzlibrary{shapes.geometric}
\usetikzlibrary{decorations.markings}
\usetikzlibrary{decorations.pathmorphing}
\usepackage{pstricks}
\def\ben{\begin{equation}}
\def\een{\end{equation}}
\def\bena{\begin{eqnarray}}
\def\eena{\end{eqnarray}}
\def\non{\nonumber}
\newcommand{\J}{{\mathcal J}}

\renewcommand{\ell}{l}
\newcommand{\sr}{\sqrt{3}}

\newcommand{\half}{\frac{1}{2}}
\newcommand{\A}{{\mathscr A}_{\rm h}}
\newcommand{\mz}{\mathbb{Z}}
\newcommand{\Q}{{\mathcal Q}}
\newcommand{\C}{{\mathcal C}}
\renewcommand{\J}{{\mathcal J}}
\newcommand{\B}{{\mathscr B}}

\theoremstyle{definition}
\newtheorem{thm}{Theorem}

\newtheorem{lemma}{Lemma}

\theoremstyle{definition}

\renewcommand{\H}{\mathscr{H}}
\newcommand{\M}{\mathscr{M}}
\renewcommand{\pounds}{{\mathscr L}}

\newcommand{\D}{\mbox{d}}

\newcommand{\mr}{\mathbb{R}}
\newcommand{\mc}{\mathbb{C}}
\newcommand{\I}{\mathscr{I}}

\newcommand{\V}{\mathcal{V}}

\newcommand{\T}{T}
\newcommand{\e}{{\rm e}}

\begin{document}

\title{Black hole uniqueness theorems and new thermodynamic identities in eleven dimensional supergravity}

\author{Stefan Hollands$^{1}$\thanks{\tt HollandsS@cardiff.ac.uk}
\\ \\
{\it ${}^{1}$School of Mathematics,
     Cardiff University,} \\
{\it Cardiff, United Kingdom} \\
}

\maketitle

\begin{abstract}
 We consider stationary, non-extremal black holes in 11-dimensional supergravity
  having isometry group $\mr \times U(1)^8$. We prove that such a black hole is uniquely specified by its angular momenta, its electric charges associated with the 7-cycles in the manifold, together with certain moduli and vector valued winding numbers characterizing
  the topological nature of the spacetime and group action. We furthermore establish
interesting, non-trivial, relations between the thermodynamic quantities associated with the black hole. These relations are shown to be a consequence of the hidden $E_{8(+8)}$ symmetry in this sector of the solution space, and are distinct from the usual ``Smarr-type'' formulas that can be derived from the first law of black hole mechanics. We also derive the ``physical process'' version of this first law
applicable to a general stationary black hole spacetime without any symmetry assumptions other than stationarity, allowing in particular arbitrary horizon topologies. The work terms in the first law exhibit the topology of the horizon via the intersection numbers between cycles of various dimension.
\end{abstract}

\section{Introduction}

Among the various supergravity theories, 11-dimensional
supergravity~\cite{CJS} plays a special role. It lives in the highest possible
spacetime dimension [in signature $(-,+,\dots,+$)], is related to most maximally supersymmetric lower dimensional supergravity theories via compactification and truncation, and has many intriguing
connections to the 10-dimensional superstring theories. It is therefore, obviously, of
considerable interest to map out the space of stationary black hole
solutions in 11-dimensional supergravity, subject to various interesting asymptotic `boundary' conditions, such as asymptotically flat, asymptotically Kaluza-Klein, asymptotically Anti-deSitter, etc.
Unfortunately, stated in this generality, this seems an almost intractable problem, because it includes by definition all such solutions in any compactification of the theory. Even worse, it would also
include solutions with very low amount of symmetry. There is evidence e.g. from the ``blackfold approach''~\cite{blackfold} that such solutions exist in higher dimensional gravity theories, but it seems unlikely that one will be able to write down some analytic expression for them.

For this reason, it seems reasonable to restrict oneself from the outset to more special stationary black hole solutions which are either (a) static, or (b) are ``algebraically special''
in a suitable 11-dimensional sense,  or (c) have fermionic symmetries, i.e. solutions to the appropriate ``Killing spinor'' equation
in 11-dimensional supergravity, or (d) have a considerable amount of bosonic symmetries, i.e. vector fields Lie-deriving the solution. Concerning (a), it seems plausible that one can classify all such solutions e.g. via the methods of~\cite{GIS1,GIS2,Gibbons02b,Masoodulalam2}, at least in the case of asymptotically flat boundary conditions (in the 11-dimensional sense). It is also conceivable that a modification/generalization of this method could be applied asymptotically Kaluza-Klein boundary conditions, but this remains to be seen. (b) A general notion of algebraically
special solutions in higher dimensions, based on the Weyl-tensor, has been proposed by~\cite{prague0,prague2}, and it has been
demonstrated that this notion is useful in principle to find/classify various special solutions e.g. in vacuum Einstein gravity, see e.g.~\cite{reall,prague1}. They include the near horizon limits~\cite{KLR07} of higher dimensional extremal black holes, although not black holes themselves. It seems likely that this strategy, complemented by a suitable condition onto the 4-form field strength, could be applied also to 11-dimensional supergravity, but presumably the same restrictions would apply. (c) This program
is pursued e.g. in papers~\cite{gutowski,gauntlet1,gauntlet2}. A complete classification was achieved in 5-dimensional minimal supergravity~\cite{harveygauntlett,Reall03} (see also the previous paper~\cite{tod} for a similar type of analysis in $N=2$
supergravity in 4~dimensions). The 5-dimensional minimal supergravity is, in some ways~\cite{mizo1}, a simpler cousin of 11-dimensional supergravity. However, in the latter case, the classification programme has yet to be completed.

In this paper, we will consider (d).
In some sense the most stringent, and symmetric, assumption is that the solution be invariant under the abelian group $\mr \times U(1)^8$, where the factor $\mr$ corresponds to the asymptotically timelike symmetry (stationarity). It is elementary to see that these assumptions restrict the asymptotic region of the spacetime $\M$ to be of the form\footnote{Of course the spacetime {\em need not} have this
topology globally a priori.} $\cong \mr^{s,1} \times T^{10-s}$, where the number of asymptotically large spatial dimensions, $s$, is either $s=1,2,3,4$. What makes this symmetry assumption special is that, as has been known for a long time~\cite{julia,mizo2}, the field equations for the bosonic fields then possess a large number of `hidden' symmetries, parameterized by the exceptional real Lie-group $E_{8(+8)}$. These hidden symmetries are useful in several ways:
\begin{enumerate}
\item[(i)] They make it possible, in principle, to generate new solutions from old ones, e.g. via the powerful variant of the ``inverse scattering method''~\cite{verdaguer}, suitably generalized to the present situation.
\item[(ii)] As we will demonstrate in sec.~\ref{sec:unique}, the hidden symmetries, together with other ideas, make it possible to derive a uniqueness theorem for black holes along similar lines as the classical results~\cite{Carter71,Carter72,Mazur84,robinson,Bunting83}. A naive expectation would be that these are uniquely characterized by their asymptotic quantities, i.e. mass $m$, angular momenta $J_i$, and the various charges $Q[C_7]$ associated with different 7-cycles $C_7$ in the asymptotic region. However, since this is already false in 5-dimensional pure gravity~\cite{ERLR,ER02a,ER02b}, it must necessarily be false also in 11-dimensional supergravity, since the respective solutions can be trivially lifted to ones in this theory. Nevertheless, generalizing a result of~\cite{HollandsYazadjiev08,HollandsYazadjiev08b}, we will show that one can define a collection of vector valued ``winding numbers'' $\{\underline{v}_J \in \mz^8\}$ associated with the action of $U(1)^8$, which encode the topology of $\M$, together with a collection of ``moduli'' $\{l_J \in \mr_+\}$. Furthermore, we prove that each connected component of the solution space, characterized by these data and the asymptotic quantities $m,J_i,Q[C_7]$, consists of at most one solution.
\item[(iii)] The hidden symmetries also make it possible to derive certain general, non-trivial relations between the thermodynamic quantities in the class of solutions under consideration. One example of such a formula is the generalization of the well-known ``Smarr relation''. In 4-dimensional Einstein-Maxwell theory this relation is $\frac{1}{4\pi}\kappa \A = m - 2\Omega J - \Phi Q - \Psi P$, where $\kappa, \A, \Omega, \Phi,\Psi, Q, P$ are respectively, the surface gravity, horizon area, horizon angular velocity, horizon electrostatic/magnetostatic potential and electric/magnetic charge. We will give an appropriate version of this relation for the black holes in 11-dimensional supergravity under consideration. More interestingly, there exist {\em further} non-trivial relations of this sort. In sec.~\ref{sec:mform}, we will derive, using the $E_{8(+8)}$ hidden symmetry, e.g. the formula
    \ben
0 = -\frac{1}{4} \ \epsilon^{jikmnpq} \ \Phi_{ik} \Phi_{mn} P_{pq} + 9 \ \delta^{jk} J_k - 8 \ m \Omega^j \ ,
    \een
    or
    \ben
    \begin{split}
& -(\delta^{jl}-\Omega^j \, \Omega^l)(\delta^{km}-\Omega^k \Omega^m) \ P_{lm} = -\frac{1}{4\pi} \ \Psi^{jk} \kappa \A + 2 \ \Psi^{mn} \Phi_{mn} Q^{jk} \\
&+ 4 \ \Psi^{l[k} \Psi^{j]m} P_{lm}
- 8 \ \Psi^{l[j} \Phi_{ml} Q^{k]m}
+ 4 \ \Psi^{jk} \Phi_{lm} Q^{lm}
\end{split}
    \een
    where $\Omega^j$, $\Phi_{ij}, \Psi^{ij}$ and $Q^{ij}, P_{ij}$ are appropriate generalizations of the angular velocities, electric/magnetic potentials and electric/magnetic charges associated with the various cycles in the horizon manifold\footnote{We will restrict ourselves in sec.~\ref{sec:mform} to a horizon of topology
    $S^2 \times T^7$, compare footnote~\ref{hortop}. The indices are related to the cycles in this manifold and run from $1,\dots,7$.}. We also derive similar other relations of this nature, the more detailed analysis of which we leave to another paper. We emphasize that all these relations are found using only the consequences of the hidden symmetry, and are not obtained from particular, explicit, solutions of the kind that we consider.
\end{enumerate}

Since we are dealing with thermodynamic relations in (iii), it is reasonable to also give the appropriate version of the most basic one, namely the first law of black hole mechanics. We will do this in sec.~\ref{sec:firstlaw}. Unlike in the rest of this paper, we are assuming here {\em only} that the black hole under consideration is stationary, but {\em not} that it is also invariant under $U(1)^8$.  In particular, one has, in principle, the possibility that the horizon manifold might be of a rather general type\footnote{\label{hortop} When the isometry group contains $U(1)^8$, the horizon topology is restricted to the entries in table~\ref{table1}. In the general case one only knows that the horizon manifold is of ``positive Yamabe type''~\cite{GallowaySchoen06}.}. The version of the first law for 11-dimensional supergravity---independent of any by-hand symmetry assumptions---is\footnote{Throughout the rest of the paper, we set the coefficient in front of the action to be $1$ rather than $1/16\pi$. This will result in trivial changes in the prefactors in the thermodynamic relations.}
\ben
\frac{1}{8\pi} \ \kappa \delta \A = \delta m - \sum_{i=1}^8 \Omega^i \delta J_i - \sum_{r,s} (I^{-1})^{rs} \Phi[C_r] \ \delta Q[C_s]
\een
where $C_r$ resp. $C_s \subset \H$ now run over the various 7-cycles resp. 2-cycles in the horizon submanifold, and where $I_{rs} \in \mz$ is the matrix of their intersection numbers.

Actually, as we recall, there are strictly speaking two interpretations of the first law, which are in effect different mathematical theorems. The difference between these interpretations concerns the nature of the variations for which the first law holds. In the more restricted, original, version~\cite{BardeenCarterHawking73} one is considering only variations within the space of {\em stationary} solutions. In the more general {\em physical process version}~\cite{waldgao,Wald01LR}, one considers also {\em non stationary} variations which satisfy the linearized equations of motion on the given black hole background, and which settle down, at late times, to a perturbation towards another stationary black hole. In this paper, we will demonstrate that the first law holds in this physical process sense~\footnote{We remark that~\cite{rogatko} has also derived a physical process version in Einstein-$p$-form theory, but he gives the work terms involving the various electric charges only implicitly, and not in the form above.}.

\medskip
\noindent
{\bf Notations and conventions:} Our conventions for the metric
and Riemann tensor follow those of~\cite{Wald84}. We also use
standard notations for differential forms; our conventions are recalled in appendix~\ref{app:nc}.  $a,b,\dots$ are $11$-dimensional spacetime
indices, while $i,j',I,J'$ are indices labeling the various Killing fields
that we assume. Throughout, we set the conventional prefactor of $1/16\pi$ in front of the
action equal to $1$ for simplicity.

\section{First law of black hole mechanics}
\label{sec:firstlaw}

\subsection{Covariant phase space method}
\label{sec:covphase}

The bosonic fields in eleven dimensional supergravity are a Lorentizan metric $g$ and a 3-form field\footnote{In particular, we are assuming in this section that $A$ is globally defined on $\M$. Therefore, there are no magnetic
 charges as automatically $\int_C F = \int_C \D A = 0$
for any closed 4-cycle $C$.} $A$ on an oriented 11-dimensional spacetime manifold $\M$. All
fermionic superpartners are set zero throughout the paper. The field equations for the bosonic fields follow from the Lagrange 11-form $L$, given by
\ben
L = R \ \star 1 -  2 \ F \wedge \star F - \frac{4}{3} \ A \wedge F \wedge F \ ,
\een
where $F = \D A$ is the associated field strength 4-form. In this paper, we are interested in stationary black hole solutions in this theory, and in the present section we would like to derive the physical process version of the first law of black hole mechanics. A convenient
formalism to derive such relations is the covariant phase space method of~\cite{waldzoupas}. This formalism
applies to any Lagrangian $L$ which is constructed locally and in a diffeomorphism covariant way out of a metric and tensor fields on an $n$-dimensional manifold $\M$, and their derivatives.
To save writing, we denote these collectively by $\psi$; in the above Lagrangian, $\psi \equiv (g,A)$.
The basic relations in the covariant phase space formalism are readily derived as follows. One considers 1-parameter families of field configurations $\psi_\lambda$, and writes $\delta \psi = \frac{d}{d\lambda} \psi_\lambda$ for the tangent (``variation'') at a given $\lambda$, e.g. $\lambda = 0$. For the above Lagrangian, $\delta \psi=(\delta g, \delta A)$. The variation of the Lagrange $n$-form may always be written as
\ben\label{delL}
\delta L (\psi) = E(\psi) \cdot \delta \psi + \D \theta (\psi, \delta \psi) \ ,
\een
where $E$ are the Euler-Lagrange equations, and
where $\D \theta$ corresponds to the ``partial integrations'' that one would carry out if the variation
was performed under an integral sign. Let $X$ be any vector field on $\M$. Then the ``Noether current'' is
the $(n-1)$-form defined by
\ben\label{jxdef}
\J_X(\psi) = \theta(\psi, \pounds_X \psi) - i_X L(\psi) \ ,
\een
where $\pounds_X$ is the Lie-derivative, and where $i_X$ is the operator that contracts the vector field into the
first index of the differential form. When the Euler-Lagrange equations $E=0$ hold, we have
\ben
\D \J_X(\psi) = 0 \ ,
\een
by a one line calculation using the formula $\pounds_X = i_X \D + \D i_X$ for the action of the Lie-derivative on a differential form.
Since this is an identity that holds for any $X$, one can prove~\cite{iyerwald} that there must always exist a
$(n-2)$-form $\Q_X$, called ``Noether charge'', locally constructed from the fields and their derivatives, such that $\D \Q_X = \J_X$. When the Euler-Lagrange equation do not hold, one can prove~\cite{iyerwald} that there is an $(n-1)$
form $\C_X$, locally constructed from the fields $\psi$ and their derivatives, and from $X$ but {\em not} its derivatives (so $\C_X = X^a \C_a$), such that
\ben\label{jx}
\J_X = \C_X + \D \Q_X \ .
\een
Clearly, $\C_X=0$ when the Euler-Lagrange equations hold, so $\C_X$ corresponds to the constraints of the theory.

One normally focusses on solutions and manifolds $\M$ obeying certain asymptotic conditions. A typical condition is that $\M$ contains an ``asymptotic region'' $\M_{\rm asymptotic}$
diffeomorphic to $\cong \mr^{n-1,1}$ minus some ``interior'',
and that the metric approaches the standard flat Minkowski metric $g_0$ at a suitable rate in this asymptotic region, whereas the other fields also obey corresponding suitable fall-off conditions. In the case of 11-dimensional supergravity, $\psi \to \psi_0$, where $\psi_0=(g_0, 0)$ consists of the Minkowski metric and the trivial 3-form field. This background configuration obviously has symmetries, $\pounds_X \psi_0 = 0$, consisting of the Killing vector fields of Minkowski space. These generate the asymptotic symmetry group $SO(n-1,1) \times \mr^n$. Other asymptotic conditions may also be considered.
For example, asymptotic Kaluza-Klein boundary conditions state that there is an asymptotic region of $\M$ modeled on $\mr^{s,1} \times \T^{n-s-1}$. This background carries the natural flat product metric $g_0$, which
is the direct product of an $s+1$-dimensional flat Minkowski metric and a flat metric\footnote{
Here we have a choice which of the non-diffeomorphic flat metrics on the torus we would like to choose.
The ``moduli space'' of such metrics is $SL(n-s-1, \mr)/SL(n-s-1, \mz)$, the local coordinates of which
include the sizes of the torus in the various diameters. In this paper, we will choose a {\em fixed} flat metric, but more generally, one could leave the particular choice unspecified. This would result in additional ``tension-type'' terms in the first law, as discussed e.g. in~\cite{traschen}.} on the torus $\T^{n-s-1}$, together with a suitable background 3-form field $A_0$, which is Lie-derived by the Killing fields of $g_0$. In that case, the asymptotic symmetry group
is $U(1)^{n-s-1} \times SO(s,1) \times \mr^{s+1}$. The precise definitions and asymptotic conditions in these cases are given in
appendix~\ref{app:ac}.

With each asymptotic symmetry, one can associate in a natural way a corresponding conserved quantity $H_X(\psi)$~\cite{waldzoupas} in the
following way. Let $\psi$ be solution to the Euler-Lagrange equations satisfying the asymptotic conditions,
and let $\delta \psi$ be a variation satisfying the linearized Euler-Lagrange equations (around $\psi$) in the asymptotic region of $\M$, but not necessarily in the interior. The conserved quantity associated with $X$
is defined by its variation through the formula
\ben\label{aquantity}
\delta H_X(\psi) = \int_{\infty} \delta \Q_X(\psi) - i_X \theta(\psi, \delta\psi)  \ ,
\een
together with the requirement $H_X(\psi_0)=0$. The notation ``$\int_\infty$'' means the following. Let $\Sigma$ be a  $(n-1)$-dimensional submanifold which in the asymptotic region approaches a suitable reference surface. For example in the case of asymptotically flat boundary conditions, with Minkowski background $g_0 = -\D t^2 + \D x_1^2 + \dots + \D x_{10}^2$, $\Sigma$ is asymptotically a $t=$ constant surface in the Minkowski background. Then we take an increasing
sequence of surfaces $\Sigma \supset C_i \cong S^{n-2}$ which smoothly approach infinity. We then evaluate the above surface integrals
$\int_{C_i}$ and take the limit as $i \to \infty$. The precise asymptotic conditions should ensure that the limit exists. The terminology ``conserved quantity'' refers to the fact that
the quantity $H_X$ does not depend on the value of $t$.  The choice $X =\partial/\partial t$ corresponds to $m$, the ``ADM mass'', the choice $X = x_i \ \partial/\partial x_j - x_j \ \partial/\partial x_i$ corresponds to the ``ADM angular momentum'' in the $ij$-plane, the choice $X = \partial/\partial x_i$ the ``ADM linear momentum''
in the $i$-direction, etc. In the asymptotically Kaluza-Klein case, where $g_0 =
-\D t^2 + \sum \D x_i^2 + \sum \D \phi_j^2$, the surfaces would be $C_i \cong S^{s-1} \times T^{n-s-1}$. We then also have the vector fields $X = \partial/\partial \phi_i$ along the generators of the torus $\T^{n-s-1}$ and corresponding conserved charges.

We now apply Stokes' theorem to eq.~\eqref{aquantity}, to obtain\footnote{\label{orientations}
Here and throughout the rest of the paper, the orientation of $\Sigma$
is fixed by the $n-1$ form defined as $\epsilon_{a_1...a_n}=
-n \ t_{[a_1} \epsilon_{a_2...a_n]}$, where $t^a$ is the future directed
timelike normal to $\Sigma$. An orientation on $\partial \Sigma$ is
fixed by $\epsilon_{a_1...a_{n-1}} = +(n-1) \ r_{[a_1} \epsilon_{a_1...a_{n-2}]}$, where $r^a$ is the spacelike normal
to $\partial \Sigma$ pointing towards the interior of $\Sigma$.
}
\ben\label{eq1}
\delta H_X = \int_\Sigma \D(\delta \Q_X - i_X \theta) + \int_{\partial \Sigma} \delta \Q_X - i_X \theta
\een
where $\partial \Sigma$ denotes any interior boundary or other asymptotic ends of $\Sigma$--if none of those are present, that term is simply set equal to zero. In this paper, we will have in mind the situation where $\Sigma$ is
a surface stretching between a cross section $\B=\partial \Sigma$ of the event horizon of a black hole and infinity. We next use
eq.~\eqref{jx} to replace
\ben\label{dqx}
\D \delta \Q_X(\psi) = \delta \J_X(\psi) - \delta \C_X(\psi) \ ,
\een
and we also use the formula~\cite{iyerwald}
\ben\label{jxom}
\delta \J_X(\psi) = \omega(\psi, \delta \psi, \pounds_X \psi) + \D \ i_X \theta(\psi, \delta \psi) \ ,
\een
where $\omega$ is the symplectic current $(n-1)$ form which depends on a pair of variations via
\ben
\omega(\psi; \delta_1 \psi, \delta_2 \psi) = \delta_1 \theta(\psi, \delta_2 \psi) -
\delta_2 \theta(\psi, \delta_1 \psi) - \theta(\psi, (\delta_1\delta_2-\delta_2\delta_1)\psi)\ .
\een
Now suppose that $X$ Lie-derives the solution $\psi$, i.e. is in particular a Killing field of the metric.
Then $\omega(\psi, \delta \psi, \pounds_X \psi)=0$, and using eqs.~\eqref{jxom},~\eqref{dqx} in eq.~\eqref{eq1}
gives
\ben\label{hxrel}
\delta H_X = - \int_\Sigma \delta \C_X + \int_{\partial \Sigma} \delta \Q_X - i_X \theta \ .
\een
This equation holds whenever $\psi$ is a solution to the Euler-Lagrange equations
satisfying the asymptotic conditions, which is Lie-derived by $X$, and for any variation
$\delta \psi$ satisfying the asymptotic conditions, and satisfying the linearized Euler-Lagrange equations near infinity (not necessarily the interior). The relation~\eqref{hxrel} will be the basis for the derivation of the first law of black hole mechanics in the next subsection. There, we will also use that, under the same conditions,
\ben
\D \delta \C_X(\psi) = \D \delta \J_X(\psi) - \D^2 \Q_X(\psi) =
\D \omega(\psi; \delta \psi, \pounds_X \psi) - \D^2 i_X \theta(\psi, \delta \psi) = 0 \ ,
\een
i.e. $\star \delta \C_X$ is a conserved current.

In the case of 11-dimensional supergravity, we find in appendix~\ref{app:nc} that the Noether charge respectively
constraints are concretely given by
\ben\label{noesugra}
\begin{split}
\Q_X &= -\star \D X - 4 \ i_X A \wedge q + \frac{4}{3} \ i_X A \wedge A \wedge F \ , \\
\C_X &= 2 \ \star f_X + 4 \ i_X A \wedge \star j \ .
\end{split}
\een
Here, we have identified $X$ (not necessarily a Killing field)
in the first term on the right side of $\Q_X$ with a 1-form, and we
have introduced the ``electric'' charge density 7-form $q$ by
\ben\label{qdef}
q = \star F +  F \wedge A \ .
\een
Furthermore, $f_X$, a 1-form, is obtained by contracting the Euler-Lagrange equation for the metric $g$ into $X$; concretely
$f_X = (G_{ab} - T_{ab}) X^a \ \D x^b$, where $T_{ab}$ is the stress
tensor, see~\eqref{tabdef}. Also, $j$, an 8-form, is the Euler-Lagrange equation for $A$. The explicit form of $j$ is given in eq.~\eqref{tjdef}; in fact, $j$ may
also be written as
\ben\label{qdef1}
\star j = \D q \ \ .
\een
$j$ is interpreted as the ``electric'' current density and $q$ as the charge density.
$f_X$ is physically interpreted as minus the flux vector of non-gravitational energy
across the horizon as seen by an observer following the flow lines of $X$.
Of course, when the Euler-Lagrange equations hold, $j = 0 = f_X$.

\subsection{Derivation of first law}
\label{sec:bhsetup}

After these preliminaries, we now derive the ``physical process version'' of the first law of black hole mechanics for 11-dimensional supergravity.
We consider solutions $(\M, g, A)$ representing a stationary black hole, satisfying either asymptotically flat or
Kaluza-Klein boundary conditions. The asymptotically timelike Killing field is denoted by $\partial/\partial t$, so $\pounds_{\partial/\partial t} g = 0 =
\pounds_{\partial/\partial t} A$. In the asymptotic region, $t$ is equal to the time-coordinate in an asymptotically Cartesian
coordinate system. We will only be concerned with the exterior of the black hole,
also called the ``domain of outer communication'', and defined more precisely by
\ben
\M_{\rm exterior} = I^-\left( \M_{\rm asymptotic} \right) \cap I^+ \left( \M_{\rm asymptotic} \right)
\een
where we mean the causal past/future of the asymptotic region. In the following we will usually write simply $\M$
again for the exterior. The future and past event horizon are then the boundary components $\partial \M =
\H^+ \cup \H^-$ lying respectively to the future/past of the asymptotic region. By construction, they are
smooth null surfaces, which may be connected (single black hole), or disconnected (multiple black holes).
For definiteness, we will restrict ourselves to single black hole spacetimes, although all of our arguments will equally apply to multiple black holes as well with trivial modifications. The situation is illustrated by the following Penrose diagram of the higher dimensional Schwarzschild/black string spacetime. In this diagram, $\M_{\rm exterior}$ is the region
shaded in blue.

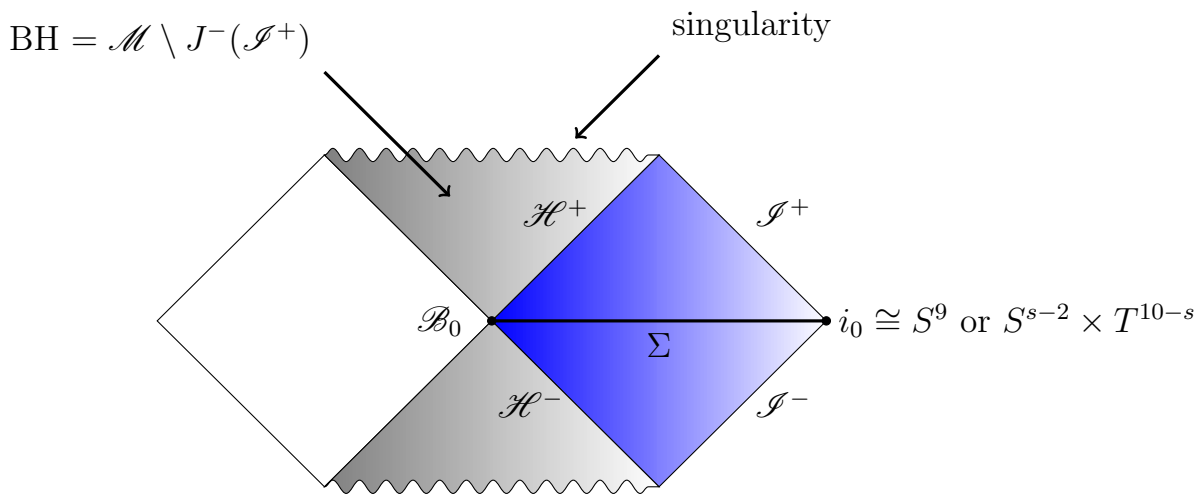
\begin{center}
\begin{tikzpicture}[scale=1.1, transform shape]
\shade[left color=blue] (0,0) -- (2,2) -- (4,0)  -- (2,-2) -- (0,0);
\draw (0,0) -- (-2,2) -- (-4,0)  -- (-2,-2) -- (0,0);
\draw (4,0) node[right]{$i_0 \cong S^{9} \ {\rm or} \ S^{s-2} \times T^{10-s}$};
\draw (0,0) -- (2,2) -- node[above right] (c) {$\I^+$}  (4,0) -- node[right] (a) {$\I^-$} (2,-2) -- (0,0);
\shade[left color=gray] (0,0) -- (-2,2) decorate[decoration=snake] {-- (2,2)} --  (0,0);
\draw (0,0) -- (-2,2) decorate[decoration=snake] {-- (2,2)} -- (0,0);
\shade[left color=gray] (0,0) -- (-2,-2) decorate[decoration=snake] {-- (2,-2)} -- (0,0);
\draw (0,0) -- (-2,-2) decorate[decoration=snake] {-- (2,-2)} -- (0,0);
\draw[->, very thick] (2,3.2) node[above right]{singularity} -- (1, 2.2);
\draw[very thick, black] (0,0) -- (4,0);
\draw (2,0) node[below]{$\Sigma$};
\filldraw (0,0) circle (.05cm);
\filldraw (4,0) circle (.05cm);
\draw[->, very thick] (-2,3) node[above left] {BH $=\M \setminus J^{-}(\mathscr{I}^{+})$} -- (-.5,1.5);
\draw (1.3,1.3) node[black,above,left]{$\H^+$};
\draw (1,-1) node[black,below,left]{$\H^-$};
\draw(-0.2,0) node[black, left]{$\B_0$};
\end{tikzpicture}
\end{center}
As is common, we restrict ourselves to the consideration of metrics
$g$ which are smooth everywhere, including an open neighborhood of the horizon $\H$. The same is also required for the field strength $F$.
However the potential, $A$, while required to be smooth away from $\H$, is allowed to be singular on $\H$; we only demand that the pull-back of $A$ to $\H$ be smooth away from the bifurcation surface $\B_0$, and that the pull-back to $\B_0$ be smooth\footnote{\label{gaofootnote}
The nature of this requirement can be illustrated in Einstein-Maxwell theory~\cite{gao}. The 1-form field $A$ in the Reissner-Nordstr\" om solution is given by $A=-(Q/r) \ \D t$,
which in Kruskal-type coordinates is $A=-(Q/2\kappa r) (U^{-1} \D U - V^{-1} \D V)$, where $\H^+=\{V=0\}, \H^-=\{U=0\}$. Clearly, the
restriction to either $\H^\pm$ is singular, but the pull-back is smooth away from the bifurcations surface $U=0=V$. A gauge could be adapted so
that $A$ becomes smooth near $\H$, but then it would either no longer be Lie-derived
by $\partial/\partial t$, or it would not decay to zero near infinity, as required by our asymptotic conditions.}.

The restriction of $\partial/\partial t$ to the event horizon $\H = \H^+ \cup \H^-$ may either point along the null generators, or not. In the first case, the black hole is said to be non-rotating; otherwise it is said to be rotating. (This notion of a rotating horizon is
logically distinct from whether the angular momenta vanish or not.) If the black hole spacetime is rotating, asymptotically flat, non-extremal, globally hyperbolic, and analytic, then the ``rigidity theorem''~\cite{HIW07,MI08,HI09} states that\footnote{See especially
\cite{HI09} for the treatment of actions with Chern-Simons type terms.} there exist $N \ge 1$ further vector fields $\xi_i$ which commute, $[\xi_i, \xi_j] = 0 = [\xi_i, \partial/\partial t]$, which Lie-derive the fields,
$\pounds_{\xi_i} g = 0 = \pounds_{\xi_i} A$, which have $2\pi$ periodic flows, and such that
\ben\label{kdef}
K = \frac{\partial}{\partial t} + \Omega^1 \ \xi_1 + \cdots + \Omega^N \ \xi_N
\een
is a Killing field which is tangent to the generators of the horizon $\H$. The constants $\Omega^i$ are referred
to as the ``angular velocities'' of the horizon. For a non-rotating black hole, $\Omega^i=0$, and $K=\partial/\partial t$. One can
show~\cite{Wald84} that the surface gravity, $\kappa$, defined by
\ben\label{kappadef}
\nabla_K K = \kappa \ K
\een
is constant (and positive) over $\H$. Since the Killing fields $\xi_i$ must belong to the 5-dimensional Cartan subalgebra of the Lie algebra of the asymptotic symmetry group $SO(10,1) \times \mr^{11}$, it is clear that $N \le 5$ in the asymptotically flat case. In the asymptotically Kaluza-Klein case, or for non-analytic stationary solutions--if these should exist--we do not have as yet an analogue of the rigidity theorem, so will simply assume the existence of the additional Killing fields $\xi_i$. Note that in this case, the asymptotic symmetry group is $U(1)^{10-s} \times SO(s,1) \times \mr^{s+1}$, whose Cartan subalgebra has dimension $10-\lfloor \frac{s}{2} \rfloor$, i.e. it is larger.

We now come to the derivation of the ``physical process'' version of the first law. We take a 10-dimensional surface $\Sigma_0$ in~\eqref{hxrel} going between a cross section $\B_0$ of the horizon to infinity as indicated in the above figure. We also take $X=K$ in~\eqref{hxrel}, and use that $H_{\partial/\partial t}=m, H_{\xi_i}=-J_i$ are the mass resp. angular momenta. This gives us
\ben
\delta m-\Omega^i \delta J_i =
-\int_{\Sigma_0} \delta \C_K + \int_{\B_0} (\delta \Q_K - i_K \theta) \ .
\een\label{balance}
As eq.~\eqref{hxrel}, this equation will hold if the
variation $(\delta g, \delta A)$ satisfies the linearized Euler-Lagrange equations near infinity. We assume this for the rest of the section. Furthermore, we make the following
{\em standing assumptions} in our derivation of the first law in this section:
 \begin{enumerate}
 \item[(i)] The variation $(\delta g, \delta A)$ vanishes in an open neighborhood of $\B_0$.
 \item[(ii)] The non-gravitational part of the stress-energy,
$t_{ab}=G_{ab}-T_{ab}$, and the non-``electro-magnetic'' part of the current, $j^{abc}$ [cf. eqs.~\eqref{tjdef}], have compact support on a later surface $\Sigma_1$ as shown in the next figure, and $(\delta g, \delta A)$ approach a perturbation to another stationary black hole at a sufficiently fast rate.
\end{enumerate}
The physical meaning of these requirements is that (i) the black hole is initially unperturbed near the horizon,  (ii) all matter and charge eventually fall into the black hole, and the perturbed black hole settles down to another stationary black hole.

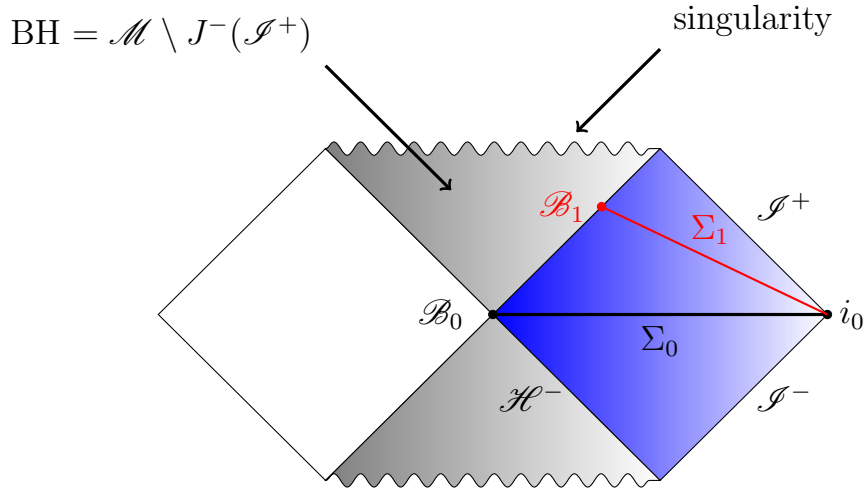
\begin{center}
\begin{tikzpicture}[scale=1.1, transform shape]
\shade[left color=blue] (0,0) -- (2,2) -- (4,0)  -- (2,-2) -- (0,0);
\draw (0,0) -- (-2,2) -- (-4,0)  -- (-2,-2) -- (0,0);
\draw (4,0) node[right]{$i_0$};
\draw (0,0) -- (2,2) -- node[above right] (c) {$\I^+$}  (4,0) -- node[right] (a) {$\I^-$} (2,-2) -- (0,0);
\shade[left color=gray] (0,0) -- (-2,2) decorate[decoration=snake] {-- (2,2)} --  (0,0);
\draw (0,0) -- (-2,2) decorate[decoration=snake] {-- (2,2)} -- (0,0);
\shade[left color=gray] (0,0) -- (-2,-2) decorate[decoration=snake] {-- (2,-2)} -- (0,0);
\draw (0,0) -- (-2,-2) decorate[decoration=snake] {-- (2,-2)} -- (0,0);
\draw[->, very thick] (2,3.2) node[above right]{singularity} -- (1, 2.2);
\draw[very thick, black] (0,0) -- (4,0);
\draw (2,0) node[below]{$\Sigma_0$};
\filldraw (0,0) circle (.05cm);
\filldraw (4,0) circle (.05cm);
\filldraw[color=red] (1.3,1.3) circle (.05cm);
\draw[->, very thick] (-2,3) node[above left] {BH $=\M \setminus J^{-}(\mathscr{I}^{+})$} -- (-.5,1.5);
\draw (1.25,1.3) node[red,left]{$\B_1$};
\draw[thick,red] (1.3,1.3) -- (4,0);
\draw (2.6,0.7) node[red,above]{$\Sigma_1$};
\draw (1,-1) node[black,below,left]{$\H^-$};
\draw(-0.2,0) node[black, left]{$\B_0$};
\end{tikzpicture}
\end{center}

Using (i), we immediately see that the second term on the right side of~\eqref{balance} is zero.  Using
(ii) and the fact that $\D \delta \C_K = 0$ for a Killing field $K$, we can write the first term on the right side as an integral over $\H^+$. Thus,
\ben
\label{balance1}
\begin{split}
\delta m - \sum_{i=1}^N \Omega_i \ \delta J_i &= -\int_{\H^+} \delta \C_K =
-2 \int_{\H^+} \star \delta f_K - 4 \int_{\H^+} i_K A \wedge \delta (\star j) + \delta (i_K A) \wedge \star j \\
&= -2 \int_{\H^+} \star \delta f_K - 4 \int_{\H^+} i_K A \wedge \D \delta q \ .
\end{split}
\een
In the  second step we have used the concrete expression for the constraints
in 11-dimensional supergravity, see eqs.~\eqref{noesugra}, whereas in the third step we used that $\star j = \D q=0$ for the electromagnetic current~\eqref{qdef} of the background solution. We next evaluate the terms on the right side. First, as we will argue momentarily, the
pull back of the 2-form $i_K A$ to $\H^+$ is closed, $\D i_K A = 0$. Therefore, the second term on the right side
of \eqref{balance1} can be written as\footnote{The choice of orientations was specified in footnote~\ref{orientations}.}
\ben\label{int}
\begin{split}
\int_{\H^+} i_K A \wedge \D \delta q &= \int_{\H^+} \D(i_K A \wedge \delta q)\\
&= -\int_{\B_0} i_K A \wedge \delta q + \int_{\B_1} i_K A \wedge \delta q \
\end{split}
\een
via Stokes' theorem, where $\B_1$ is a  cross section of the horizon
as indicated in the figure, which is ``later'' than the support of $\delta j$. Since
the variation vanishes by assumption (i) near $\B_0$, the first integral
on the right side is zero. To write the second integral in more recognizable
form, we choose basis of cycles
\ben
C_r \in H_7(\B_1, \mz) \ , \quad C_s \in H_2(\B_1, \mz) \ , \quad
I_{rs} := \# (C_r  \cap C_s) \in \mz \ ,
\een
in the 9-dimensional horizon cross section $\B_1$.
$I_{rs}$ denotes the matrix of intersection numbers, i.e.
the number of intersection points counted with $\pm$ signs
determined by the relative orientations. Then setting
\ben\label{Qdef}
\begin{split}
\Phi[C_r] &= -\int_{C_r} i_K A \ , \quad r = 1, \dots, \dim H_2(\B_1, \mz) \\
Q[C_s] &= 4\int_{C_s} q \ , \quad s = 1, \dots, \dim H_7(\B_1, \mz)
\end{split}
\een
and using de-Rahm's theorem, the integral~\eqref{int} becomes
\ben\label{int1}
-4 \int_{\H^+} i_K A \wedge \delta j = \sum_{r,s} (I^{-1})^{rs} \Phi[C_r] \delta Q[C_s] \ .
\een
The number $\Phi[C_r]$ is interpreted as the ``electrostatic potential'' of the horizon associated with the 2-cycle $C_r$ of the horizon cross section,
whereas $Q[C_s]$ is interpreted as the ``electric'' charge associated with the 7-cycle $C_s$. It remains to be shown that the pull back of $\D i_K A$ to $\H^+$ vanishes. We have
\ben
\D i_K A = -i_K \D A + \pounds_K A = -i_K F \ ,
\een
so we need to show that $i_K F = 0$ when pulled back to $\H^+$.  Let $k$ be a vector field tangent to
affinely parameterized null geodesic generators of $\H^+$. It is evidently
proportional to $K$ at every point of $\H^+$. By eq.~\eqref{kappadef}, if $U$ is the affine parameter, so that $k=\partial/\partial U$ in suitable coordinates,
the relation is in fact $K = \kappa U \ \partial/\partial U$. The Raychaudhuri equation states
that
\ben\label{raych}
\begin{split}
\frac{\D}{\D U} \vartheta &= - \frac{1}{9} \vartheta^2 -
\sigma_{ab} \sigma^{ab} - R_{ab} k^a k^b \\
& = - \frac{1}{9} \ \vartheta^2 -
\sigma_{ab} \sigma^{ab} - T_{ab} k^a k^b
\end{split}
\een
where $T_{ab}$ is the stress tensor of the 3-form field, see eq.~\eqref{tabdef}, and where
in the second line we used Einstein's equation. Now,
for a stationary black hole one finds by the same argument as used in
the area theorem that $\vartheta$, the expansion of the geodesic
congruence generated by $k$, vanishes on $\H^+$. Consequently,
since all three terms on the right side are non-positive, we must
have $\vartheta = 0 = T_{ab} k^a k^b$ on $\H^+$. By eq.~\eqref{tabdef}, this gives
$F_{acde} F_b{}^{cde} k^a k^b = 0$ on $\H^+$, which in turn implies that $i_k F = k \wedge \alpha$ for some 2-form $\alpha$, where $k$ has been identified with a 1-form via $g$. Viewed as a 1-form, $k$ has vanishing pull-back to $\H^+$, therefore so has $i_k F$, which completes the argument.

We now evaluate the first integral on the right side in our balance equation~\eqref{balance1}. Using the definition
$f_X=(G_{ab}-T_{ab}) X^a \ \D x^b$, his may be written alternatively as
\ben
-2 \int_{\H^+} \star \delta f_K = 2 \kappa \int_{\H^+} U \ \delta \bigg( R_{ab} k^a k^b - T_{ab} k^a k^b \bigg) \epsilon
\een
where $\epsilon$ is the positively oriented 10-dimensional volume element on
$\H^+$ defined by $k \wedge \epsilon = -\star 1$. For the variation of
the Ricci tensor component $R_{ab} k^a k^b$, we obtain from the variation of
the Raychaudhuri equation
\ben
\delta (R_{ab} k^a k^b) = \delta R_{ab} k^a k^b + 2 R_{ab} k^a \delta k^b =
-\frac{\D}{\D U} \delta \vartheta  +2 T_{ab} k^a \delta k^b \ .
\een
We may assume that we are in a gauge such that $\delta k^a$ is proportional to
$k^a$. Then from $T_{ab} k^a k^b = 0$ on the horizon, it follows that $T_{ab} k^a \delta k^b = 0$ on the horizon. Using this, and
the formula~\eqref{tabdef} for $T_{ab}$, we similarly find
\ben
\begin{split}
\delta (T_{ab} k^a k^b) &= \frac{2}{3} \ k^a \delta F_{abcd} k^e F_e{}^{bcd} - \delta g^{ef} k^a F_{acde} k^b F_{bf}{}^{cd}\\
&=\frac{2}{3} \ k^a \delta F_{abcd}  k^{[b} \alpha^{cd]} - \delta g^{ef} k_{[c} \alpha_{de]} k^b F_{bf}{}^{cd} \\
&= 0 \ ,
\end{split}
\een
using $\delta g_{ab} k^a  \propto k_b$.
Thus, we have shown that the first integral
in the balance equation~\eqref{balance1} is
\ben\label{int2}
 -2 \int_{\H^+} \star \delta f_K
= -2 \kappa \int_{\H^+} U \ \frac{\D}{\D U} \delta \vartheta \ \epsilon
= 2 \kappa \delta \A \ ,
\een
where the second equality follows from the calculation in sec.~6.2 of
\cite{waldbook}, and where $\delta \A$ is the variation of the area of
a horizon cross section at asymptotically late times\footnote{To make the
integral over $\H^+$ converge, we are assuming at this stage our assumption (ii) that the perturbation settles down to a perturbation to another stationary black holes at a sufficiently fast rate.}. Combining eqs.~\eqref{balance1}, \eqref{int1}, \eqref{int2}, we get
\ben
2 \kappa \delta \A = \delta m - \Omega^i \delta J_i - \sum_{r,s} (I^{-1})^{rs} \Phi[C_r] \ \delta Q[C_s] \ .
\een
This is the desired first law of black hole mechanics in 11-dimensional
supergravity. Evidently, it has the same general form as the first law in Einstein-Maxwell theory, but the detailed expression of the term involving ``electric'' potentials and charges depends on the topology of the 9-dimensional compact horizon cross section in question. It enters via the matrix of intersection numbers $I_{rs}$ between the 2-cycles and 7-cycles inside the
horizon cross section.

\section{Black hole uniqueness theorem}
\label{sec:unique}

We now prove a uniqueness theorem for stationary asymptotically Kaluza-Klein black holes having
$8$ commuting rotational vector fields $\xi_1, \dots, \xi_8$ which Lie-derive $g$ and $A$, and which also commute with the action
of the asymptotically time-like Killing field $\partial/\partial t$. Hence, from now we assume
that the isometry group is $\mr \times U(1)^8$. This is consistent with an asymptotic
region of the form $\mr^{s,1} \times T^{10-s}$ for $s=1,2,3,4$ large spatial dimensions.
For definiteness, we will stick to the case $s=4$.

\subsection{Structure of orbit space, Weyl-Papapetrou form}
\label{sec:wp}

As a first step towards a uniqueness theorem, we have to better understand the global nature
of the spacetime $\M$, its topology, and the global nature of the action of $U(1)^8$. Furthermore,
as in the vacuum theory, it is essential to use the symmetries and field equations in order to construct particularly useful coordinates. First, to set up some notation, we introduce the 8 dimensional Gram matrix of the rotational Killing fields, denoted by
\ben\label{fdef}
f_{ij} = g(\xi_i, \xi_j) \ .
\een
We say that a point $P \in \M$ is on an ``axis'' if there is a linear combination
\ben
\sum_{i=1}^8 v^i \ \xi_i \bigg|_P = 0 \ .
\een
It is not difficult to see~\cite{HollandsYazadjiev08b} that $\underline{v} = (v^1, \dots, v^8)$ must be in $\mz^8$ up to some overall rescaling, and we fix that rescaling by the requirement $g.c.d.(v^1, \dots, v^8)=1$, where $g.c.d.$ is the greatest common divisor. The particular linear combination will in general depend on the point $P$.
The first major step is the following theorem:

\begin{thm}\label{thm:wp}
(``Weyl-Papapetrou-form'')
The metric can be brought into Weyl-Papapetrou form~\eqref{wp}
away from the horizon $\H$ and any axis of rotation.
\ben\label{wp}
g = - \frac{r^2 \, \D t^2}{\det f} + \e^{-\nu} (\D r^2 + \D z^2)
+
f_{ij}( \D \phi^i + w^i \, \D \tau)(\D \phi^j + w^j \, \D \tau) \, ,
\een
where $\phi^i$ are $2\pi$-periodic coordinates such that the rotational Killing fields $\xi_1, \dots, \xi_8$
take the form $\xi_i=\partial/\partial \phi^i$, and where $t$ is a coordinate such that
the timelike Killing field takes the form $\partial/\partial t$. In other words,
the metric functions $f_{ij}, w^i, \nu$ are independent of $t, \phi^1, \dots, \phi^8$,
and only depend on $z \in \mr, r > 0$.
\end{thm}

{\em Proof:} The proof of this statement is given for the vacuum theory in $n$ dimensions in~\cite{HollandsYazadjiev08b} relying e.g. on
global results such as topological censorship~\cite{Galloway99,CGS09,fws} and results on
 spaces with torus actions generalizing those of~\cite{orlik1,orlik2}; here we will only outline the (minor) modifications that have to be made in the case of 11-dimensional supergravity. The proof consists of essentially four ingredients:
(a) The distribution of subspaces $({\rm span}(\partial/\partial t,\xi_1,\dots,\xi_8))^\perp \subset T\M$ is
locally integrable for a solution to the equations of motion in 11-dimensional supergravity. This expresses that the metric~\eqref{wp}
has no ``cross terms'' between $(r,z)$ and $(t,\phi^1,\dots,\phi^8)$. (b) The orbit space
$\hat \M=\M/G$, with $G = \mr \times U(1)^8$ the isometry group, is diffeomorphic to an upper half plane. This upper half plane is parameterized in~\eqref{wp} by the coordinates $(r>0, z)$. In particular, implicit here is the--very far from obvious!--statement that 2-dimensional orbit space e.g. {\em cannot have any conifold points, nor holes, nor handles.}
(c) The function $r$ on $\M$ defined by
\ben\label{rdef}
r^2 = -{\rm det}
\left(
\begin{matrix}
g(\partial_t,\partial_t) & g(\partial_t,\xi_j)\\
g(\xi_i,\partial_t) & g(\xi_i, \xi_j)
\end{matrix}
\right)
\een
is globally defined on $\M$, except at the axis and the horizon. In particular, the right side is positive everywhere on $\M$, except at these places, where it is zero. This expresses that the span of $\partial/\partial t,\xi_1, \dots, \xi_8$ is everywhere a timelike distribution of 9-dimensional
subspaces of $T\M$, except at the horizon or the axis. (d) The function
$r$ is a harmonic function on the orbit space $\hat \M = \M/G$. This enables one to define
$z$ as the conjugate harmonic function on $\hat \M$, so $\hat \M$ is parameterized by
$\hat \M = \{(r,z) \mid r >0\}$ away from the axis and the horizon. Since $(r,z)$ is
a harmonically conjugate pair, the induced metric on the orbit space is $\e^{\nu}(\D r^2 + \D z^2)$ for some conformal factor $\e^{\nu}$.
Statements (a)--(d) are equivalent to the statement that the metric
takes the form~\eqref{wp}.

The claims (a),(d) are local in nature and follow from the equations of motion. By contrast,
the claims (b),(c) are global in nature. Their proof involves the equations of motion, but also global techniques from topology. The global properties (b),(c) in particular imply that the coordinate system can be constructed {\em globally}, away from the horizon and the axis.

The proof of (a) is standard for $n$-dimensional vacuum general relativity and 4-dimensional
Einstein-Maxwell theory. Here we give the proof in the case of 11-dimensional supergravity. For definiteness, we repeat the statement:
\begin{lemma}\label{lem:frob}
The distribution of subspaces $({\rm span}(\partial/\partial t,\xi_1,\dots,\xi_8))^\perp \subset T\M$ is
locally integrable for a solution to the equations of motion in 11-dimensional supergravity.
\end{lemma}
{\em Proof:}
Let us denote the Killing fields collectively as $\xi_I, I=0,\dots,8$, with $\xi_0=\partial/\partial t$,
and $\xi_I, I=1, \dots, 8$ denoting the remaining rotational Killing fields. In view of the
``differential forms version'' of Frobenius' theorem, we must prove that $\D \xi_I =
\alpha^J{}_I \wedge \xi_J$ for some 1-forms $\alpha^I{}_J$, or what is the same
\ben\label{frobthm}
0= \xi_0 \wedge \dots \wedge \xi_8 \wedge \D \xi_I
\een
for any $I$. Here, we have identified the Killing vectors with 1-forms via the metirc.
Consider the Noether-charge 9-form $\Q_{\xi_I}$. Since $\xi_I$ Lie-derives the fields $g,A$,
it follows from eq.~\eqref{jxdef} and $\D \Q_{\xi_I} = \J_{\xi_I}$ that
\ben
\D \Q_{\xi_I} = -i_{\xi_I} L \ .
\een
We now contract all Killing fields $\xi_0, \dots, \xi_8$ into both sides of this equation.
The Killing field $\xi_I$ gets contracted twice into the form $L$ on the right side, so we get zero:
\ben\label{dqn}
\begin{split}
0 =& \ \ i_{\xi_0} \cdots i_{\xi_8} \D \Q_{\xi_I} \\
=& \ \ i_{\xi_0} \cdots i_{\xi_8} \D \left(
-\star \D \xi_I - 4 \ i_{\xi_I} A \wedge q + \frac{4}{3} \ i_{\xi_I} A \wedge A \wedge F
\right) \\
=& \D i_{\xi_0} \cdots i_{\xi_8}(\star \D \xi_I)
+i_{\xi_0} \cdots i_{\xi_8}\bigg( -4 \ \D(i_{\xi_I} A) \wedge q  \\
& \hspace{3cm} - 4 \ i_{\xi_I} A \wedge \D q+ \frac{4}{3} \ \D(i_{\xi_I} A) \wedge A \wedge F + \frac{4}{3} \ i_{\xi_I} A \wedge F \wedge F
\bigg)
\\
=& \D [\star(\D \xi_I \wedge \xi_8 \wedge \cdots \wedge \xi_0)]
+ i_{\xi_0} \cdots i_{\xi_8} \left(4 \ i_{\xi_I} F \wedge q - \frac{4}{3} \ A \wedge i_{\xi_I} F \wedge F +
\frac{4}{3} \ i_{\xi_I} A \wedge F \wedge F
\right) \ .
\end{split}
\een
Here, we have used the concrete expression for the Noether charge~\eqref{noesugra},
and we have used repeatedly the fact that $i_{\xi_J} \D + \D i_{\xi_J} = \pounds_{\xi_J}$, together with the fact that $\pounds_{\xi_J}$ annihilates any expression formed out of $g,A$,
and the $\xi_K$, by $\pounds_{\xi_J} \xi_K = [\xi_J, \xi_K]=0$. We have also used $\D A = F$
and $\D q=0$, by the equations of motion, see eq.~\eqref{qdef}. We now carry out the contractions $i_{\xi_J}$ into the expression in parenthesis on the right side. When four
$i_{\xi_J}$'s hit the 4-form $F$, we get $=0$, again by
\ben
i_{\xi_{J_1}} i_{\xi_{J_2}} i_{\xi_{J_3}} i_{\xi_{J_4}} F = -i_{\xi_{J_1}} \D (i_{\xi_{J_2}} i_{\xi_{J_3}} i_{\xi_{J_4}} A)
= -\pounds_{\xi_{J_1}}  (i_{\xi_{J_2}} i_{\xi_{J_3}} i_{\xi_{J_4}} A)=-i_{\xi_{J_1}} i_{\xi_{J_2}} i_{\xi_{J_3}} \pounds_{\xi_{J_4}} A = 0 \ .
\een
Hence, the only term on the right side of \eqref{dqn} which potentially is not zero is
the one where precisely 7 insertion operators hit the charge 7-form $q$. To see that such
a term vanishes as well, we note that, by $\D q=0$, we have
\ben
\D(i_{\xi_{J_1}} \cdots i_{\xi_{J_7}} q) = i_{\xi_{J_1}} \cdots i_{\xi_{J_7}} \D q = 0 \ ,
\een
so the scalar function $i_{\xi_{J_1}} \cdots i_{\xi_{J_7}} q$ is constant on $\M$. However,
by the fall-off conditions imposed on the field $A$ (see appendix~\ref{app:ac}), it vanishes at infinity, so it must be equal to zero. Hence, we conclude from eq.~\eqref{dqn} that the scalar function $\star(\D \xi_I \wedge \xi_8 \wedge \cdots \wedge \xi_0)$ must be constant on $\M$. Since there must be at least one point $P \in \M$ where one linear combination of the $\xi_J$'s vanishes by the orbit space theorem, see below, it follows that this quantity must in fact be zero, hence~\eqref{frobthm} follows.
\qed

A more detailed version of statement (b), which elucidates also the nature of the
action of the rotational isometry group $U(1)^8$, is given in the following theorem,
which is proved in the same way as that in~\cite{HollandsYazadjiev08b} for $n$-dimensional
vacuum general relativity:

\begin{thm}\label{thm:orb}
(``Orbit space theorem'') If one assumes the isometry group $G = \mr \times U(1)^8$, then the orbit space
$\hat \M = \M/G$ is homeomorphic to an upper half plane $\{(r,z) \mid r>0\}$. Furthermore, the
boundary $r=0$ can be thought of as divided up into a collection of intervals $(-\infty, z_1), (z_1, z_2),
\dots, (z_n, +\infty)$, each of which either represents the orbit space $\hat \H = \H/G$ of the horizon
(one interval per horizon component, if multiple horizons are present), or an axis in the spacetime
where a linear combination $\sum_i v_{J}^i \psi_i$ of the rotational Killing fields vanishes. The quantity $\underline{v}_J \in \mz^8$ is a vector associated with the $J$-th interval which necessarily has
integer entries.
For adjacent intervals $J$ and $J+1$ (not including the horizon), there is a compatibility condition stating that
the collection of minors $\mu_{kl} \in \mz, \,\, 1 \le k < l \le 8$ given by
\ben\label{Qkl}
\mu_{kl} = \left|\det \left(
\begin{matrix}
v_{J+1}^k & v_J^k\\
v_{J+1}^l & v_J^l
\end{matrix}
\right) \right| \,
\een
have greatest common divisor ${g.c.d.}\{ \mu_{kl} \}= 1$.
\end{thm}

\medskip
\noindent
{\bf Remark:}
As in the vacuum theory,
the relation between $r,z$, and the asymptotically Cartesian (spatial) coordinates $x_1, \dots, x_s$ in the asymptotically
KK-region~(see sec.~\ref{sec:bhsetup}) is
\ben\label{rzasy}
(r,z) \sim \begin{cases}
(\sqrt{x_1^2 + x_2^2}, x_3) & \text{if $s=3$}\\
(\sqrt{(x_1^2+x_2^2)(x_3^2+x_4^2)}, \half(x_1^2 + x_2^2 - x_3^2 - x_4^2)) & \text{if $s=4$}
\end{cases}
\een

Statement (c) can be proved in the same way as in vacuum general relativity, for
a proof see~\cite{Chr09}. (d) holds whenever the theory can be locally dimensionally
reduced to a certain kind of sigma model on a symmetric space, as proved in~\cite{breitgibbons}. This type of sigma model reduction is recalled for 11-dimensional supergravity in sec.~\ref{sec:sigma}.
\qed

\medskip
\noindent


Some examples of interval structures in 5-dimensional vacuum
general relativity are summarized in the following table.
\begin{center}
\begin{tabular}{|c|c|c|c|}
\hline
               & Interval Lengths & Vectors (Labels) & Horizon \\
\hline
Myers-Perry & $\infty, l_1, \infty$ & $(1,0),
(0,0), (0,1)$ & $S^3$ \\
\hline
Black Ring & $\infty, l_1, l_2, \infty$ & $(1,0),
(0,0), (1,0), (0,1)$ & $S^2 \times S^1$\\
\hline
Black Saturn & $\infty, l_1, l_2, l_3, \infty$ & $(1,0), (0,0), (0, 1), (0,0), (0,1)$ & $S^3 \ \
{\rm and} \ \ S^2 \times S^1$\\
\hline
Black String & $\infty, l_1, \infty$ & $(1,0), (0,0), (1,0)$ & $S^1 \times S^2$\\
\hline
Black Di-Ring & $\infty, l_1, l_2, l_3, l_4, \infty$ & $(1,0),(0,0),(1,0),(0,0),(1,0),(0,1)$ & $2 \cdot (S^1 \times S^2) $  \\
\hline
Orth. Di-Ring & $\infty, l_1, l_2, l_3, l_4, \infty$ & $(1,0),(0,0),(1,0),(0,1),(0,0),(0,1)$ & $2 \cdot (S^1 \times S^2)$  \\
\hline
Minkowski & $\infty, \infty$ & $(1,0), (0,1)$ & ---\\
\hline
\end{tabular}
\end{center}
In this table, the interval $(0,0)$ corresponds to a horizon.
The explicit form of the metric may be found in~\cite{MP86} (Myers-Perry),
\cite{PS06,ER02b} (Black ring), \cite{Elvang&Figueras07} (Saturn),
\cite{IguchiMishima07} (Di-Ring), and~\cite{Izumi} (Orthogonal Di-Ring). Of course, all these solutions can be lifted trivially
to solutions in 11-dimensional supergravity; the vectors $\underline{v}_J$ would be
turned into 8-dimensional vectors by filling the remaining 6 components with $0$'s.

The sequence of vectors $\underline{v}_J$ encodes the entire information about the toplogy of $\M$, and the nature of the group action. In particular, the horizon topology is specified. More precisely if $\underline{v}_{h-1}$
resp. $\underline{v}_{h+1}$ are associated with the intervals adjacent to a horizon interval $(z_h, z_{h+1})$  we can say for example the following: Let $\mu_{kl} \in \mz, 1 \le k < l \le 8$ be the integers
defined from these two vectors as in eq.~\eqref{Qkl}, and set
$p = g.c.d.(\mu_{kl})$. This parameter is related to the
different horizon topologies by the table~\ref{table1}.
\begin{figure}
\begin{center}
\begin{tabular}{c|c}
$p$ & Topology of horizon cross section \\
\hline
$0$ & $S^2 \times T^7$ \\
$\pm 1$ & $S^3 \times T^6$ \\
other & $L(p,q) \times T^6$
\end{tabular}
\caption{\label{table1}
   \small{The invariant $p= g.c.d.(\mu_{kl})$ characterizes the
   different horizon topologies.} }
\end{center}
\end{figure}
Note that the first and last vector $\underline{v}_0, \underline{v}_N$ in the above solutions is always $(1,0)$ resp. $(0,1)$. This corresponds to the fact that these 5-dimensional solutions are asymptotically flat in all 5 directions. In the case of 11-dimensional supergravity with 5 asymptotically Minkowskian dimensions, we would instead have $(1,0,0,0,0,0,0,0)$ and $(0,1,0,0,0,0,0,0)$. For a more detailed discussion of the interval structure see~\cite{HollandsYazadjiev08b,Harmark04}. This finishes our
review of the orbit space theorem, and we now explain how to actually construct the Weyl-Papapetrou  coordinates~\eqref{wp}.

\subsection{Sigma model reduction and divergence identities}
\label{sec:sigma}

It is well-known that the field content of 11-dimensional supergravity can be
reorganized into that of a gravitating sigma model into a certain coset when the theory
is dimensionally reduced from 11 dimensions to three dimensions~\cite{julia}. This formulation can, and will, be used in the proof of our black hole uniqueness theorem below, and also in sec.~\ref{sec:mform}. Therefore, we briefly review the construction, following the treatment given in~\cite{mizo2}. The formulation relies on the introduction of certain scalar potentials, and we now describe a conceptually simple way of defining these which also makes manifest their relation to the global conserved quantities of the theory, used later.

First, consider the closed ``electric'' charge 7-form $q$, see eq.~\eqref{qdef}. Contracting this into 6 out of the 8 rotational Killing fields generating the action of $U(1)^8$, we get
a 1-form, $i_{\xi_{i_6}} \cdots i_{\xi_{i_1}} q$. This 1-form is immediately seen to be closed
using the identity $i_{\xi_i} \D + \D i_{\xi_i} = \pounds_{\xi_i}$ on forms together with the
fact that $\pounds_{\xi_i}$ annihilates any tensor field that is constructed from $g,A,\xi_j$
and their covariant derivatives. Hence, at least locally, there exists a scalar function
$\chi_{i_1 \dots i_6}$ such that
\ben\label{chi1def}
\D \chi_{i_1 \dots i_6} = i_{\xi_{i_6}} \cdots i_{\xi_{i_1}} q \ .
\een
Next, consider the Noether charge 9-form $\Q_{\xi_i}$, see eq.~\eqref{noesugra}.
By exactly the same argument as in the proof of lemma~\ref{lem:frob}, we see that
the 1-form $i_{\xi_1} \cdots i_{\xi_8} \Q_{\xi_i}$ is closed. Hence, at least locally,
there exists a scalar function $\chi_i$ such that
\ben\label{chi2def}
\D \chi_i = i_{\xi_1} \cdots i_{\xi_8} \Q_{\xi_i} \ .
\een
Even though $\M$ is not simply connected, the ``twist potentials'' $\chi_i, \chi_{i_1 \dots i_6}$ are in fact defined globally. This follows from the fact that, since they
are invariant under the action of the isometry group $\mr \times U(1)^8$, the defining equations can be viewed as equations for closed 1-forms on the simply connected orbit space
$\hat \M = \{(r,z) \mid r>0\}$. A third set of scalars is defined by contracting three rotational Killing fields into the three form $A$,
\ben
A_{i_1 i_2 i_3} = i_{\xi_{i_1}} i_{\xi_{i_2}} i_{\xi_{i_3}} A \ .
\een
It turns out that the field equations for the solutions $(g,A)$ invariant under $\mr \times U(1)^8$ can be written entirely in terms of the 128 scalars $(f_{ij},\chi_j, \chi_{i_1 \dots i_8}, A_{k_1k_2k_3})$; in fact, they can be thought of as parameterizing a single field
that is valued in the coset space $E_{8(+8)}/SO(16)$ which has precisely this dimension. This construction is also useful for
us, so we review it following~\cite{mizo2}. Recall that the real\footnote{Unless stated otherwise, all Lie-groups and Lie-algebras in this paper are real, e.g. $SL(9)$ means
$SL(9, \mr)$, etc.} Lie-algebra ${\frak e}_{8(+8)}$ contains $\frak{sl}(9)$ as a subalgebra. As a vector space, it is given by
\ben\label{decomp}
\frak{e}_{8(+8)} = \frak{sl}(9) \oplus (\mr^9)^{\wedge 3} \oplus (\mr^{9*})^{\wedge 3} \ .
\een
This is also how the adjoint representation of $\frak{e}_{8(+8)}$ decomposes when
restricted to $\frak{sl}(9)$.
The corresponding 80+84+84=248 generators are $(e^I{}_J, e^{*IJK}, e_{IJK})$ where the
$e^I{}_J$ generate $\frak{sl}(9)$, and where capital Roman letters $I,J,\dots$
run from $1, \dots, 9$. The star symbol on $e^{*IJK}$ is part of the name of the generator, and does not mean any kind of conjugation or dual. Relations and other relevant basic facts about this Lie-algebra are recalled in appendix~\ref{app:e8}. Let us define the 8-bein $e_i{}^{\hat a}$ by $f_{ij} = \delta_{\hat a \hat b} e_i{}^{\hat a} e_j{}^{\hat b}$, and form the following $SL(9)$ matrix:
\ben
V = \left(
\begin{matrix}
e_i{}^{\hat a} & \det e^{-1}( \chi_i + \frac{1}{720} \ \epsilon^{j_1 \dots j_8} A_{ij_1 j_2} \chi_{j_3 \dots j_8}) \\
0 & \det e^{-1}
\end{matrix}
\right) \ .
\een
We also define the $\frak{e}_{8(+8)}$ valued function $v$ by
\ben
v =  e^{IJK} A_{IJK} + \frac{1}{360} \ e^*_{IJK} \epsilon^{IJKLMNPQR} \chi_{LMNPQR}
\een
where $A_{IJK}=2 \sqrt{3} \ A_{ijk}$ when $I=i,J=j,K=k$ are between $1, \dots, 8$, and
zero otherwise, as well as similarly $\chi_{LMNPQR} = 2\sqrt{3} \ \chi_{lmnpqr}$
if all indices are between $1, \dots, 8$ and zero otherwise. We can now form the
248-dimensional matrix $\V$ valued in the adjoint representation of the group
$E_{8(+8)}$ by\footnote{It can be shown that $v$ is nilpotent, ${\rm ad}(v)^5 = 0$,
so the exponential is in fact a polynomial in the components of $v$ of degree 4.}
\ben\label{Vdef}
\V = \exp({\rm ad}(v)) \ {\rm Ad}(V) \ .
\een
Here, in the last expression, $V \in SL(9)$ has been viewed as an element of $E_{8(+8)}$
in accordance with the decomposition~\eqref{decomp}. As is common, ``${\rm Ad}$''
refers to the adjoint representation of the group on its Lie-algebra, whereas ``${\rm ad}$''
to the adjoint representation of the Lie-algebra on itself. Let $\tau$ be the involution
on $E_{8(+8)}$ given in appendix~\ref{app:e8}, and define the matrix
\ben\label{Mdef}
M = \V \ \tau(\V)^{-1} \ .
\een
Then it is shown in~\cite{mizo2} that the equations of motion for $M$ on the orbit space
$\hat \M = \{(r,z) \mid r>0\}$ derived from the action
\ben
I = \int_{\hat \M} r \  |M^{-1} \D M|^2_{k,\hat g}
\ \D \hat v
\een
are exactly equivalent to those that can be derived for the
 128 scalars $(f_{ij}, \chi_j, \chi_{i_1 \dots i_8}, A_{k_1k_2k_3})$ from the 11-dimensional
supergravity Lagrangian\footnote{To make the identification with the quantities used in~\cite{mizo2}, we should identify their potentials $\psi_i$ resp. $\varphi^{ij}$ with
$\chi_i + \frac{1}{6!} \epsilon^{j_1 \dots j_8} A_{ij_1 j_2} \chi_{j_3 \dots j_8}$
resp. $\frac{1}{6!}\epsilon^{ijk_1 \dots k_6} \chi_{k_1 \dots k_6}$.}.
In the above formula, $k(X,Y)=-{\rm Tr}({\rm ad}(X){\rm ad}(Y))$ indicates the Cartan Killing form
where $\D \hat v$ is the integration element of the orbit space metric $\hat g= \D r^2 + \D z^2$.

\medskip
\noindent

The action $I$ can be viewed as that of of non-linear sigma model in the symmetric space
$E_{8(+8)}/SO(16)$, as follows. First, recall that a symmetric space is defined generally as
a triple $(G,H,\tau)$, where $G$ is a Lie-group with involution\footnote{An involution
is a homorphism $g \mapsto \tau(g)$ of $G$ such that $\tau^2=id$ for all
$g \in G$.} $\tau$, and
$H$ is a Lie-subgroup of $G$ satisfying $G^\tau_0 \subset H \subset G_{}^\tau$, with a superscript $\tau$ denoting the elements invariant under $\tau$, and with the subscript $0$
denoting the connected component of the identity. Given a symmetric space, one can define
the principal $H$-bundle $G \to X = G/H$ (right cosets). This principal fibre bundle has a global section defined by $X \owns gH \mapsto g\tau(g)^{-1} \in G$. If $G$ is semi-simple (as in our example $G=E_{8(+8)}$) then it carries a natural metric from the Cartan-Killing form
on $\frak{g}$, and the pull-back of this metric via the above global section then gives
a metric $\mathcal G$ on $X=G/H$. It is a general theorem about symmetric spaces that if
$G$ is simply connected and non-compact, and $H$ maximally compact, then (a) that metric $\mathcal G$ is Riemannian, and (b)
it has negative sectional curvature, see thm.~3.1 of~\cite{Helgason}. By negative curvature, one means more
precisely the following. Let ${\mathcal R}_{ABCD}$ be the Riemann tensor of ${\mathcal G}_{AB}$. The Riemann tensor is
always anti-symmetric in $AB$ and $CD$, and symmetric under the exchange of $AB$ with $CD$. Thus, it can be viewed as a symmetric, bilinear form ${\rm Riem}: \wedge^2 T_\sigma X \times \wedge^2 T_\sigma X \to \mr$ in each tangent tangent space of $X$, where $\sigma = gH$
denotes an element of $X$. We say that $(X, {\mathcal G})$ has negative sectional curvature (or simply, is ``negatively curved'') if this
bilinear form only has negative eigenvalues. In other words, there is a $c > 0$ such that, for any anti-symmetric 2-tensor $\omega$ we have ${\rm Riem}(\omega, \omega) \le -c \|\omega \|^2$, or in components,
\ben
{\mathcal R}_{ABCD} \omega^{AB} \omega^{CD} \le -c \ {\mathcal G}_{AC} {\mathcal G}_{BD} \omega^{AB} \omega^{CD} \ .
\een
We are precisely in this case, if $G=E_{8(+8)}$, if $\tau$
is defined as in appendix~\ref{app:e8}, and if $H=SO(16)=G^\tau_0$.
In fact, writing as above $\V \in E_{8(+8)}$ and $M=\V \tau (\V)^{-1}$, the metric on $X$ is by construction equal
to ${\mathcal G} = -{\rm Tr}(M^{-1} \D M \otimes M^{-1} \D M)$, and
$I$ is consequently equal to the action of a non-linear sigma model from the upper
half plane $\hat \M$, taking values in $X$.

Suppose now that $\sigma_\lambda: \hat \M \to X$ is a family of solutions to the sigma model equations of motion, where $\sigma_\lambda = \V_\lambda \cdot SO(16)$ is the right coset of the matrix $\V_\lambda \in E_{8(+8)}$ of the solution given above in eq.~\eqref{Vdef}. Let $\delta \sigma_\lambda = \frac{\partial}{\partial \lambda} \sigma_\lambda: \hat \M \to \sigma_\lambda^* TX$ be the linearization at any fixed $\lambda$, interpreted as the infinitesimal displacement of the 2-dimensional ``worldsheet'' in $X$ swept out by $\sigma_\lambda$. Then, one can derive from the sigma-model field equation the following equation for
$\delta \sigma_\lambda \equiv \delta \sigma$:
\ben\label{variation}
\frac{1}{r}\hat \nabla (r\hat \nabla \delta \sigma^A) = {\mathcal R}^A{}_{BCD}(\sigma) \ (\D \sigma^B) \cdot (\D \sigma^D) \ \delta \sigma^C \ .
\een
Here $\hat \nabla$ is the natural derivative operator in the bundle $\sigma^* TX \to \hat \M$ that is inherited from
the derivative operator of the metric $\mathcal G$ on $X$.
This equation may be viewed as the generalization of the ``geodesic deviation equation'' in $X$ from curves to
surfaces.

Now assume that $[0,1] \owns \lambda \mapsto \sigma_\lambda$ connects two given solutions $\sigma_0, \sigma_1$ of the sigma model equations coming from two corresponding matrix functions $\V_0, \V_1$. Define the
scalar valued function $S$ on $\hat \M$ by
\ben\label{Sdef}
S(x) := \int_0^1 \bigg( {\mathcal G}_{AB}(\sigma_\lambda) \delta \sigma_\lambda^A \delta \sigma_\lambda^B \bigg)(x)  \ \D \lambda \ge 0 \ .
\een
Then it is straightforward to derive from eq.~\eqref{variation} the following formula
\bena
&&\frac{1}{r} \hat \nabla (r \hat \nabla  S) \\
&=& \int_0^1 \D \lambda \left( {\mathcal G}_{AB}(\sigma_\lambda) \ \hat \nabla \delta \sigma_\lambda^A \cdot \hat \nabla \delta \sigma_\lambda^B - {\mathcal R}_{ABCD}(\sigma_\lambda) \ (\delta \sigma_\lambda^A \D \sigma_\lambda^C ) \cdot (
\delta \sigma_\lambda^B \D \sigma_\lambda^D)
\right) \non\\
&\ge& \int_0^1 \D \lambda \left( {\mathcal G}_{AB} \ \hat\nabla \delta \sigma_\lambda^A \cdot \hat\nabla \delta \sigma^B_\lambda + c \ {\mathcal G}_{AB} {\mathcal G}_{CD} \ (\delta \sigma_\lambda^{[A} \D \sigma_\lambda^{C]} ) \cdot (
\delta \sigma_\lambda^{[B} \D \sigma_\lambda^{D]})
\right) \ge 0 \ , \non
\eena
where to get the key $\ge 0$ relations we have used (a) that the target space is Riemannian, and (b) that it
is negatively curved. A slightly different way of writing this differential inequality, which is useful in the next section, is to define a fictitious $\mr^3$ parameterized by
$(x = r \sin \varphi, y = r \cos \varphi, z)$, and to view $S$, rather than as a function of
$\hat \M = \{(r,z) \mid r>0\}$, as an axially symmetric function on this $\mr^3 \setminus \{z-{\rm axis}\}$. Then the differential inequality is simply
\ben\label{subharm}
\left(
\frac{\partial^2}{\partial x^2} + \frac{\partial^2}{\partial y^2} +
\frac{\partial^2}{\partial z^2}
\right) S(x,y,z) \ge 0 \ ,
\een
in other words $S \ge 0$, may be viewed as a sub-harmonic function on the fictitious $\mr^3$ minus
the $z$-axis.
We will make use of this property in the next section.

\subsection{Uniqueness proof}

Suppose that we are given two non-extremal, stationary,
black hole solutions $(\M_0, g_0, A_0)$ respectively $(\M_1, g_1, A_1)$, asymptotic
to $\mr^{4,1} \times T^6$,
both of which are invariant under invariant under the action of the group $\mr \times U(1)^8$, with corresponding commuting Killing fields $\xi_0 = \partial/\partial t, \xi_1, \dots, \xi_8$. In this section we investigate under which conditions these
solutions must in fact be isometric. Obviously, we can assign to the solutions the
ADM-conserved quantities $m, J_1, \dots, J_8$, and electric charges~\eqref{Qdef} associated with any 7-cycles in the respective spacetimes. Furthermore, by the orbit space theorem, we can assign to each of the solutions the interval structure, i.e. the invariant interval lengths
$l_J \in \mr_+$, and the corresponding vectors of winding numbers $\underline{v}_J \in \mz^8$.
We claim:

\begin{thm}
Consider two non-extremal, single horizon, stationary black hole solutions $(\M_0, g_0, A_0)$ respectively $(\M_1, g_1, A_1)$ whose isometry group is $\mr \times U(1)^8$, which are asymptotically Kaluza-Klein,
and which satisfy the other global regularity conditions (a)--(e) stated in appendix~\eqref{app:ac}. Suppose the solutions
can be connected by a (differentiable) 1-parameter family of such solutions
$(\M_\lambda,g_\lambda, A_\lambda)$ having the same ADM conserved quantities $m,J_i$,
having the same electric charge associated with any 7-cycle, and having the same interval
lengths $\{l_J\}$. Then $(g_0, A_0)$ is equal to $(g_1, A_1)$ up to a diffeomorphism.
\end{thm}

\noindent
{\bf Remarks:} 1) The same method of proof also applies in the case of multiple non-degenerate horizons. In that case, we may form for each horizon cross section $\B_k$ the
quantities $J_i[\B_k], i=1,\dots,8$ defined by eq.~\eqref{komar}, with $C=\B_k$ there. These numbers, interpreted as the angular momentum for the $k$-th black hole, must be required to be the same for both solutions for all $k$. \\
2) One should be able to prove that the solutions must agree even if
one does not know, a priori, that they can be connected. To get this stronger result,
one should use the distance function in Mazur's identity~\cite{Mazur84} instead of the distance function $S$ as in our proof. We leave this to a graduate student who is not afraid of the tiresome, although straightforward, $E_{8(+8)}$-algebra. A uniqueness theorem for 5-dimensional minimal supergravity using Mazur's identity was given in~\cite{tomizawa1,tomizawa2}. By contrast to our theorem, these authors however assume by hand a particularly simple interval structure (and hence topology).

\medskip
\noindent

{\em Proof:} It is clear that the spacetime manifolds $\M_\lambda$ must be diffeomorphic and that the action of the group $\mr \times U(1)^8$ must be equivalent for all $\lambda$. The orbit space theorem characterizes the action of $U(1)^8$. Therefore, the two solutions must
have the same winding numbers $\{ \underline{v}_J \}$ described in the orbit space theorem. Of course, the orbit space theorem makes no statement about the dynamical fields, so we do not know yet what is the relation between $(g_\lambda, A_\lambda)$ for different $\lambda$. For this 1-parameter family, consider the function $S: \mr^3 \setminus
 \{z{\rm -axis}\} \to \mr$ defined above in eq.~\eqref{Sdef}. (Recall that the coordinates
 $(x,y,z)$ of this $\mr^3$ are related to the coordinates of the orbit space $\hat \M =
 \{(r,z) \mid r>0\}$ by $x = r \cos \varphi, y=r \sin \varphi$, where $\varphi$ is an angle that does not have any straightforward relation to the coordinates on the spacetime $\M$.)
 The function $S$ is sub-harmonic, eq.~\eqref{subharm}, and non-negative. We wish to apply ``Weinstein's lemma''~\cite{weinstein} to $S$.

\begin{lemma}
(``Weinstein's lemma'')
Let $S(x,y,z) \ge 0$ be a bounded function on $\mr^3$ which is continuous on $\mr^3 \setminus \{z{\rm -axis}\}$ and which is is a solution to
$(\partial_x^2 + \partial_y^2 + \partial_z^2) S \ge 0$, in the distributional sense. Then $S=const$.
\end{lemma}

To apply this lemma, we need to verify that our $S$ is uniformly bounded, including near infinity and the  $z$-axis. At this stage we need the assumptions about the mass, interval structure, angular momenta, and charges. First, let us write down explicitly the
function $S = \int_0^1 \D \lambda s_\lambda$. We put $\frac{\D}{\D \lambda}(g_\lambda, A_\lambda) \equiv (\delta A, \delta g)$, $(g_\lambda, A_\lambda) \equiv (g,A)$. Then
relying on calculations in~\cite{mizo1}, we find, with $\sigma = \V \cdot SO(16)$ as above:
\ben
\begin{split}
&s_\lambda = {\mathcal G}_{AB}(\sigma_\lambda) \delta \sigma_\lambda^A \delta \sigma_\lambda^B = \\
&240 \ \Bigg(
\frac{1}{4} \ f^{ik} f^{jl} \delta f_{ij} \delta f_{kl} + (\delta \log \det f)^2 + \frac{1}{3}
\ f^{il} f^{jm} f^{kn} \delta A_{ijk} \delta A_{lmn} \\
& + 10 \ f^{n_1m_1} \dots f^{n_6m_6}( \frac{1}{60} \ \delta \chi_{n_1\dots n_6} - \frac{1}{3} \ \delta A_{[n_1n_2n_3} A_{n_4n_5n_6]})(\frac{1}{60} \ \delta \chi_{m_1\dots m_6} - \frac{1}{3} \ \delta A_{[m_1m_2m_3} A_{m_4m_5m_6]}) \\
& + \frac{1}{2} \frac{1}{\det f} \ f^{ij} (
\delta \chi_j + \frac{1}{54} \ \epsilon^{k_1\dots k_8} A_{jk_1k_2} A_{k_3k_4k_5} \delta A_{k_6k_7k_8} + \frac{1}{360} \ \epsilon^{k_1 \dots k_8}
A_{jk_1k_2} \delta \chi_{k_3 \dots k_8}
)\\
& \hspace{2.5cm}(
\delta \chi_i + \frac{1}{54} \ \epsilon^{l_1\dots l_8} A_{il_1l_2} A_{l_3l_4l_5} \delta A_{l_6l_7l_8} + \frac{1}{360} \ \epsilon^{l_1 \dots l_8}
A_{il_1l_2} \delta \chi_{l_3 \dots l_8})
\Bigg) \ .
\end{split}
\label{sla}
\een
Clearly, $S$ is a sum of squares, as it has to be. To show that it is uniformly bounded, we thus need to show that each of the terms is uniformly bounded individually on $\mr^3 \setminus
\{z{\rm -axis}\}$. By construction, $S$ is smooth, so we need to make sure that it does not blow up anywhere near the $z$-axis, nor at infinity. The behavior near infinity can be controlled in a straightforward manner because the asymptotic behavior of the fields $(g,A)$,
hence $(\delta g, \delta A)$ is known. Maybe the only point to note here is that we have to use the precise relation between the coordinates $r,z$ and the asymptotically Cartesian coordinates in the spacetime $\M$, which is given above in eq.~\eqref{rzasy}. The asymptotic behaviors of $A_{ijk}, f_{ij}$, viewed as functions on $\mr^3$, then follow immediately from the behavior of the fields in the original asymptotically Cartesian coordinates. Likewise, it is straightforward to determine the behaviors of the potentials $\chi_i, \chi_{ij\dots k}$, by
integrating up the defining relations~\eqref{chi2def} resp.~\eqref{chi1def}. If this is done, then it is found that $s_\lambda$, hence $S$, is uniformly bounded for large $r^2 + z^2$.
The details of this arguments are very similar to those given in~\cite{HollandsYazadjiev08b} in the vacuum theory, so we do not give them here.

The more tricky part is to control the behavior of $s_\lambda$, hence $S$, near the $z$-axis
in $\mr^3$. Recall that the $z$-axis is the union of intervals $[z_J, z_{J+1}]$, where each
interval represents points in the original spacetime $\M$ that are either on a horizon, or
which are on an axis of rotation.

\medskip
\noindent
{\bf Axis of rotation:} On a point $P \in \M$ on an axis of rotation, an integer linear combination $v_J^i \xi_i |_P = 0$, or equivalently $f_{ij} v_J^j |_{(r,z)}= 0$, where
$r,z$ are the Weyl-Papapetrou coordinates of $P$. Furthermore,
unless $P$ is a ``turning point'', $v^i_J$ spans the null space of $f_{ij}$. On the other hand, if  $P$ is a turning point, then it lies on the boundary, say $z=z_J$, of an interval, and the  null space of $f_{ij}$ is two-dimensional and spanned by $v_J^i, v_{J-1}^i$. Owing to the
condition on subsequent vectors $v_J^i, v_{J-1}^i$ stated in the orbit space theorem, one can see~\cite{HollandsYazadjiev08b} that there exists a $SL(8,\mz)$ matrix $B_i{}^j$ such that
\ben
B_i{}^j v_J^i = (1,0,0,0,0,0,0) \ , \quad
B_i{}^j v_{J-1}^i = (0,1,0,0,0,0,0,0) \ .
\een
By redefining the rotational Killing fields of the spacetime if necessary by this matrix as $\xi_i \to B_i{}^j \xi_j$ globally (corresponding to the conjugation of the action of $U(1)^8$ on $\M$ by an inner automorphism),
we may assume that the vectors $v_J^i, v_{J-1}^i$ take the above simple form. Then it follows
that the components of $f_{ij}, f^{ij}$ have the following behavior near a point $(r,z)$
where $z$ is in an interval $(z_J, z_{J+1})$ representing an axis of rotation:
\ben\label{fbe}
f_{ij} = \begin{cases}
O(r^2) & \text{if $i=1$ or $j=1$,}\\
O(1) & \text{otherwise,}
\end{cases}
\qquad
f^{ij} = \begin{cases}
O(r^{-2}) & \text{if $i=j=1$,}\\
O(1) & \text{otherwise.}
\end{cases}
\een
Likewise, since $\xi_1$ vanishes on our axis of rotation, we have
\ben\label{abe}
A_{ijk} = \begin{cases}
O(r^2) & \text{if $i=1$, or $j=1$, or $k=1$}\\
O(1) & \text{otherwise.}
\end{cases}
\een
We also need the behavior of the other potentials $\chi_i, \chi_{ij \dots k}$ for
$(r,z)$ approaching the interval $(z_J, z_{J+1})$. This is slightly more tricky
and requires using the information about the asymptotic charges. Consider first $\chi_i$,
with defining relation~\eqref{chi2def}. It follows from the fact that some linear combination
of the $\xi_i$ vanishes on each axis of rotation, that the left side of this relation is $=0$, hence $\D \chi_i = 0$
on any axis of rotation, i.e. on any interval $(z_J, z_{J+1})$ marked in red in the following figure. Hence, $\chi_i=const.$ on the red line, but not the horizon, marked in blue.
Since we are free to add constants to $\chi_i$, we may e.g. assume that $\chi_i=\pm const$ to the left/right of the horizon interval. This constant may be
computed as follows. Consider a curve $\hat \gamma$ as in the figure, going from a point $z'$ the right of the horizon to another point $z$ to the left of the horizon.

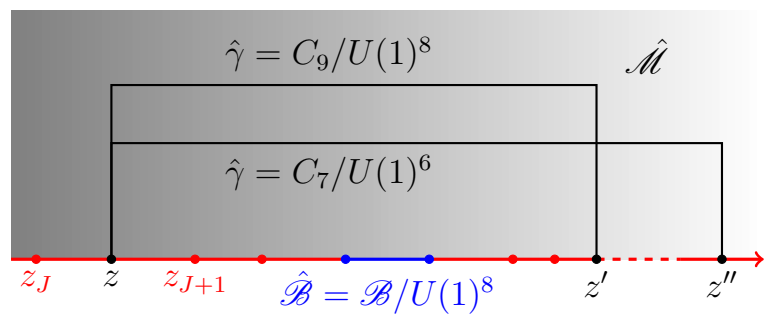
\begin{center}
\begin{tikzpicture}[scale=1.1, transform shape]
\shade[left color=gray] (-4,0) -- (5,0) -- (5,3)  -- (-4,3) -- (-4,0);
\draw[very thick,color=red] (1,0) -- (3,0);
\draw[very thick,color=red,dashed] (3,0) -- (4,0);
\draw[->,very thick,color=red] (4,0) -- (5,0);
\draw[very thick,color=red] (-4,0) -- (0,0);
\draw[very thick,color=blue] (0,0) -- (1,0);
\draw (.5,0) node[below,color=blue]{$\hat \B = \B/U(1)^8$};
\draw (3.6,2.8) node[below]{$\hat \M$};
\draw (3,0) node[below]{$z'$};
\draw (-2.8,0) node[below]{$z$};
\draw[black,thick] (3,0) -- (3,2.1) -- (-2.8,2.1) -- (-2.8,0);
\draw[black,thick] (4.5,0) -- (4.5,1.4) -- (-2.8,1.4) -- (-2.8,0);
\draw (-0.2,1.4) node[below]{$\hat \gamma=C_7/U(1)^6$};
\draw[black] (4.5,0) node[below]{$z''$};
\draw (-0.2,2.1) node[above]{$\hat \gamma=C_9/U(1)^8$};
\filldraw[color=blue] (1,0) circle (.05cm);
\filldraw[color=blue] (0,0) circle (.05cm);
\filldraw[color=red] (-1.8,0) circle (.05cm);
\draw[color=red] (-1.8,0) node[below]{$z_{J+1}$};
\filldraw[color=red] (-3.7,0) circle (.05cm);
\draw[color=red] (-3.7,0) node[below]{$z_{J}$};
\filldraw[color=red] (-1,0) circle (.05cm);
\filldraw[black] (4.5,0) circle (.05cm);
\filldraw[color=red] (2,0) circle (.05cm);
\filldraw[color=red] (2.5,0) circle (.05cm);
\filldraw[color=black] (-2.8,0) circle (.05cm);
\filldraw[color=black] (3,0) circle (.05cm);
\end{tikzpicture}
\end{center}

It is the equivalence class of
a 9-dimensional cycle $C_9$ in $\M$ under the quotient by $U(1)^8$, i.e. $\hat \gamma =
C_9/U(1)^8$. (The topology of this $C_9$ depends on the integer vectors associated with the
intervals that $z,z'$ are in, respectively, see table~\ref{table1}.) Now we compute
\ben
\begin{split}
(2\pi)^8 \ \chi_i \bigg|_{(z,r=0)}^{(z',r=0)} & = (2\pi)^8 \ \int_{\hat \gamma=C_9/U(1)^8} \D \chi_i \\
& = \int_{C_9} \D \chi_i \wedge \D\phi^1 \wedge \cdots \wedge \D \phi^8 \\
& = \int_{C_9} \Q_{\xi_i} \\
&= \int_{C_9} \Q_{\xi_i} - i_{\xi_i} \theta = H_{\xi_i} = -J_i \ ,
\end{split}
\een
where $\Q_{\xi_i}$ is the Noether charge, and where we used~\eqref{chi2def}, together with the
definition of the ADM-conserved quantity $J_i$ associated with the Killing fields $\xi_i =
\partial/\partial \phi^i$. Hence, we conclude that
\ben
\chi_i(z,r) =
\pm \frac{1}{2}(2\pi)^8 \ J_i + O(r^2)
\een
Here $\pm$ is chosen depending on whether $z$ is to the left/right
of the horizon. Consequently, because we are assuming that $J_i$ are the same for our solutions, we have
\ben\label{chi1be}
\delta \chi_i = O(r^2) \ ,
\een
near any interval representing an axis, i.e. the red lines.
Consider next the potentials $\chi_{i_1 \dots i_6}$,
with defining relation~\eqref{chi1def}. It follows from the fact that $\xi_1$ vanishes on
on $(z_J, z_{J+1})$ that the left side of this relation is $=0$ when one of the
$i_k$'s is equal to 1. Therefore $\D \chi_{1i_1 \dots i_5} = 0$ on the interval $(z_J, z_{J+1})$, i.e. $\chi_{1i_1 \dots i_5} = 0$ must be constant there. We now set this
constant in relation to the electric charge of an appropriate 7-cycle, just as we did with the
angular momenta before. Let us add constants to $\chi_{i_1 \dots i_6}$ in such a way
that these potentials tend to 0 as $r=0, z \to + \infty$. Then, consider a curve $\hat \gamma$
connecting $z''$ at infinity and $z \in (z_J, z_{J+1})$ as in the following figure. This curve
corresponds to the image of a 7-cycle under the quotient $\hat \gamma = C_7/U(1)^6$,
where the subgroup $U(1)^6$ is generated by $\xi_1, \dots, \xi_6$. We now get
\ben
\begin{split}
(2\pi)^6 \ \chi_{12 \dots 6} \ \bigg|_{(z',r=0)}^{(z,r=0)} & = (2\pi)^6 \ \int_{\hat \gamma=C_7/U(1)^6} \D \chi_{12 \dots 6} \\
& = \int_{C_7} \D \chi_{12 \dots 6} \wedge \D\phi^1 \wedge \cdots \wedge \D \phi^6 \\
& = \int_{C_7} q = Q[C_7] \ .
\end{split}
\een
Since the electric charges associated with any cycle are assumed to be the same for the solutions, we find for all $i,j,\dots,k=1,\dots,8$:
\ben\label{chi2be}
\delta \chi_{ij \dots k} =
\begin{cases}
O(r^2) & \text{if $i=1$, or $j=1$, or $\dots$, or $k=1$}\\
O(1) & \text{otherwise.}
\end{cases}
\een
This concludes our analysis of the potentials near an axis of rotation. Combining
eqs.~\eqref{fbe}, \eqref{abe}, \eqref{chi1be}, \eqref{chi2be} with \eqref{sla}, one immediately sees that
$s_\lambda$, hence $S$, is bounded near the interval $(z_J, z_{J+1})$. Of course, this
analysis applies to the interior of any boundary interval representing an axis. It may also be shown by the same type of analysis as in~\cite{HollandsYazadjiev08b} that the same holds true at the turning points, i.e. where two intervals
representing an axis intersect.

\medskip
\noindent
{\bf Horizon:} The analysis of the behavior of $s_\lambda$, hence $S$, on the horizon
interval $(z_h, z_{h+1})$ is not problematical, since there is no linear combination of
the $\xi_i's$ which vanishes there. Hence, the Gram matrix of these Killing fields,
$f_{ij}$ must be non-singular on the horizon interval. Hence, by contrast to
the intervals associated with an axis of rotation, all fields appearing in $s_\lambda$
have a continuous limit as the interval is approached, and $s_\lambda$ hence remains bounded.
Some care is however required at the endpoints $z_h, z_{h+1}$. Here one has to be alert that
near these points, the Weyl-Papapetrou coordinates give a rather distorted picture of the
spacetime geometry, as they are not smooth there. Consequently, one has to rule out the possibility that $s_\lambda$, hence $S$, might have very direction dependent--and possibly singular--limits as $z_h,z_{h+1}$ is approached. The comprehensive, rather tedious, analysis of this point was given in~\cite{HollandsYazadjiev08b} in the vacuum case, where it was shown that $s_\lambda$ remains bounded no matter from what direction these points are approached. That analysis also carries over, with very few modifications, to the present case, so we omit it here.

\medskip
\noindent

In summary, we have shown that $S$ is uniformly bounded on $\mr^3$, including crucially the
$z$-axis and infinity. Hence, by Weinstein's lemma, $S = const.$, and since $S$ decays in some directions, $S=0$, and hence $s_\lambda=0$. Hence, $\delta \chi_i, \delta \chi_{ij \dots k},
\delta A_{ijk}, \delta f_{ij}$ all vanish (for any $\lambda$), and hence the fields $\chi_i, \chi_{ij \dots k},
A_{ijk}, f_{ij}$ associated with the two solutions $(g_0, A_0)$ and $(g_1, A_1)$ agree. One must still show that the solutions agree themselves. This is seen as follows.
First, it follows from the duality relation~\eqref{chi1def} that the
functions $w_j = f_{ij}w^i$ in the Weyl-Papapetrou form~\eqref{wp} can be obtained in terms of the scalar
potentials parameterizing the matrix $M$ as
\ben
\begin{split}
&\D w_j = \\
&-\frac{r}{\det f} \ \hat \star \bigg( \D \chi_j + \frac{1}{54} \ \epsilon^{k_1\dots k_8} A_{jk_1k_2} A_{k_3k_4k_5} \D A_{k_6k_7k_8} + \frac{1}{360} \ \epsilon^{k_1 \dots k_8}
A_{jk_1k_2} \D \chi_{k_3 \dots k_8} \bigg) \ ,
\end{split}
\een
Then it follows that $w_i$ must agree for the two solutions. The function $\nu$ in~\eqref{wp} can be recovered from the matrix $M$ [see~\eqref{Mdef}] using eq.~3.32 in~\cite{breitgibbons}, so $g_0=g_1$.
It also follows from the duality relation~\eqref{chi2def} that the components
$A_{0ij}$ can be obtained then by integrating
\ben
\begin{split}
\D A_{0ij} &= -\frac{1}{720} \ \frac{r}{\det f} f_{im} f_{jn} \epsilon^{mnk_1 \dots k_6} \ \hat \star(\D \chi_{k_1 \dots k_6}) + w^k \D A_{kmn} \\
&+ \frac{1}{36} \ \frac{r}{\det f} \ f_{im} f_{jn} \epsilon^{mni_1 \dots i_6}
\ \hat \star(\D A_{i_4 i_5 i_6}) A_{i_1 i_2 i_3} \ ,
\end{split}
\een
where the subscript ``$0$'' here refers to the contraction with the Killing field
$\xi_0=\partial/\partial t$.
Hence, these components of the $A$-field agree for the two solutions.
The 1-form components $A_{ij}, A_{0i}$ and 2-form components $A_0, A_i$ are likewise seen to agree using the
field equations for the $A$-field. This completes the proof.
\qed

\section{Mass formulas}
\label{sec:mform}

\subsection{Komar-type expressions for $m,J_i$ and Smarr formula}

Above, we have given a general formula for the ADM-type conserved quantity $H_X$ associated with any asymptotic symmetry $X$, see eq.~\eqref{aquantity}. Although that expression
can be integrated to give a completely explicit formula very similar to the
standard expressions in the case of 4-dimensional vacuum general relativity~\cite{Wald84},
we will use here another, ``Komar-type'', formula. This formula is less general because it holds only if the asymptotic symmetry
$X$ in question is an actual symmetry of the solution under consideration, i.e. Lie-derives
$(g,A)$. As in the previous section, we assume that the spacetime is stationary, with asymptotically time-like Killing field $\xi_0=\partial/\partial t$, together with $N$
commuting rotational Killing fields $\xi_i$ as in eq.~\eqref{kdef}.

\begin{lemma}\label{lem:Komar}
Let $X$ be any vector field which Lie-derives the solution $(g,A)$. Then the following
9-form $\alpha_X$ is closed:
\ben
\D \alpha_X = 0 \ , \qquad \alpha_X = - \star \D X - \frac{8}{3} \ i_X A \wedge q - \frac{4}{3} \ A \wedge \star(F \wedge X) \ .
\een
Furthermore, the mass $m=H_{\partial/\partial t}$ and angular momenta $J_i = -H_{\xi_i}$ can be expressed as
\ben\label{komar}
m = \frac{9}{8} \int_C \alpha_{\partial/\partial t} \ , \qquad J_i = -\int_C \alpha_{\xi_i} \ ,
\een
where, $C$ is any 9-dimensional cycle\footnote{The orientations are chosen as in footnote~\ref{orientations}.} cobordant to spatial infinity.
\end{lemma}
{\bf Remark:} The cycle $C$ may be chosen so that $\xi_i$ is tangent. Then the formula for
the angular momentum reduces to $J_i = \int_C \star \D \xi_i$, which is the standard ``Komar-type'' expression in the case of vacuum general relativity. Similarly, if $A=0$,
then the formula for the mass becomes $m = -\frac{9}{8} \int_C \star \D \xi_0$, which is the Komar
mass formula for 11-dimensional vacuum general relativity. The factor of $-9/8$ is the analogue in 11-dimensions of the factor $-2$ discrepancy between the Komar mass and angular momentum expression in 4 dimensions~\cite{Wald84}.

Before we give a proof of this lemma, let us apply it to get a ``Smarr-type'' formula
in 11-dimensional supergravity. Consider the Killing vector field $K$ as in eq.~\eqref{kdef}, which
is tangent to the null generators of the horizon $\H$. By Stokes theorem, since
$\alpha_K$ is a closed 9-form, we get $\int_\B \alpha_K = \int_\infty \alpha_K$, where $\B$
is the bifurcation surface of the horizon, and where $\infty$ is a cross section at infinity.
Using the lemma, this gives
\ben\label{72}
\int_\B \alpha_K = \frac{8}{9} m - \sum_i \Omega^i J_i \ .
\een
We now evaluate the integral on the left side. A standard calculation~\cite{Wald84} using
eq.~\eqref{kappadef} gives $\int_\B \star \D K = -2 \kappa \A$, where $\A$ is the horizon area. Also, note that $i_K A = -\Phi$ is the electrostatic `potential' on the horizon,
which we have already shown in sec.~\ref{sec:firstlaw} to be a closed 2-form on $\H$, hence $\B$. Finally, note that
since $K$ itself vanishes on $\B$, since $A$ has vanishing pull-back\footnote{See footnote~\ref{gaofootnote}.} to
$\B$, since $K$ vanishes on $\B$, and since $\star F$ is smooth, we have $A \wedge i_K \star F = 0$ when pulled back to $\B$. Consequently, defining the charge for a 7-cycle $C_r \subset \B$ as above in~\eqref{Qdef} by $Q[C_r]$, and and the electrostatic
potential associated with a 2-cycle $C_s \subset \B$ by $\Phi[C_s]$, the left side of~\eqref{72} evaluates
to
\ben
2 \ \kappa \A + \frac{2}{3} \sum_{r,s} (I^{-1})^{rs} Q[C_r] \Phi[C_s] =
\int_\B \alpha_K \ ,
\een
where $I_{rs}$ is the intersection matrix.
Hence, the Smarr formula is
\ben\label{smarr}
9 \cdot 2 \ \kappa \A = 8\ m - 9 \ \Omega^i J_i - 6 \  (I^{-1})^{rs} Q[C_r] \Phi[C_s]
 \ .
\een
It can also be obtained directly from the first law by
considering a variation as in the proof of the lemma.

\medskip
\noindent

{\em Proof of lemma~\ref{lem:Komar}:}
 We first note that if we scale the solution
$(g,A)$ as $\phi_\lambda:=(\lambda^2 g, \lambda^3 A)$ for a constant $\lambda$, then the Lagrangian of
11-dimensional supergravity changes as $L(\phi_\lambda) = \lambda^9 L(\phi_1)$. Variation along this 1-parameter family of rescaled field configurations hence gives, using eq.~\eqref{delL}, $9 \ L = \D \theta$,
where $\theta$ is evaluated on the variation $\delta \phi=(2g,3A)$. Using the explicit form of $\theta$ given in appendix~\ref{app:nc} gives that $L$ is expressible as
$L = -\frac{4}{3} \D ( A \wedge \star F )$, whenever $(g,A)$ satisfies the equations of motion.
Now suppose additionally that a vector field $X$ Lie-derives this solution. Then from eq.~\eqref{jxdef}, together with $i_X \D + \D i_X = \pounds_X$ and the fact that $\pounds_X$
annihilates any tensor fields built from $(g,A)$:
\ben\label{ax}
\D \Q_X = -i_X L = \frac{4}{3} \ i_X \D (A \wedge \star F) = -\frac{4}{3} \ \D \bigg[ i_X A \wedge \star F - A \wedge \star( F \wedge X) \bigg] \ .
\een
Write the difference between the two sides as $\D \alpha_X = 0$.
Using the explicit form $\Q_X$, see eq.~\eqref{noesugra}, we find that $\alpha_X$
is given by the formula claimed in the lemma.

We now show the formulas for $m, J_i$. First consider the angular momenta. Let us choose a cross section at infinity such that $\xi_i$ is tangent. Then the term involving $i_{\xi_i}$
in the formula~\eqref{aquantity} vanishes, and we get $\delta H_{\xi_i} = \delta \int_\infty \Q_{\xi_i}$, hence $-J_i = H_{\xi_i} = \int_\infty \Q_{\xi_i}$. But if $\xi_i$ is tangent to the cross section at infinity, the last expression is also equal to $\int_\infty \Q_{\xi_i} =
\int_{\infty} \alpha_{\xi_i}$, and because $\alpha_{\xi_i}$ is closed, we may deform the cross
section at infinity in the last expression to any other 9-cycle $C$. This proves the formula
for $J_i$ in the lemma.

To prove the formula for $m$, consider
a 1-parameter family of diffeomorphisms which
acts in the asymptotic region as a dilatation $f_\lambda: x^\mu
\mapsto \lambda^{-1} x^\mu$, where $x^\mu$ are the asymptotically Cartesian coordinates at infinity--of course this is not an isometry in non-trivial cases. Let $\psi_\lambda^{} = f^*_\lambda \phi_\lambda^{}$, where $\phi_\lambda$ is the rescaled solution defined above. Then this solution is again asymptotically flat in the sense described in app.~\ref{app:ac}; the dilatation ensures that the metric asymptotes to the Minkowski metric in the asymptotic region, with conformal factor equal to 1. Note that $(f_\lambda^{-1})_* \frac{\partial}{\partial t} \to \lambda^{-1} \ \frac{\partial}{\partial t}$ in the asymptotic region, hence the vector field $Y$ generating $f_\lambda$ has commutator $[Y, \frac{\partial}{\partial t}] \to \frac{\partial}{\partial t}$. We now consider the defining relation \eqref{aquantity} for $\delta H_{\partial/\partial t}$ under the variation along the family $\psi_\lambda$. On the one hand, we have, with $\psi \equiv \psi_1$:
\ben
\begin{split}
\frac{\D}{\D \lambda} \ H_{\partial/\partial t}(\psi_\lambda) \Big|_{\lambda=1}
&=
\frac{\D}{\D \lambda} \ [H_{\partial/\partial t}(f_\lambda^* \psi) + H_{\partial/\partial t}(\phi_\lambda)]_{\lambda=1} \\
&=
\frac{\D}{\D \lambda} \ [H_{(f_\lambda^{-1})_* \partial/\partial t}(\psi) + \lambda^9 H_{\partial/\partial t}(\psi)]_{\lambda=1} \\
&=\frac{\D}{\D \lambda} \ [\lambda^{-1} H_{\partial/\partial t}(\psi) + \lambda^9 H_{\partial/\partial t}(\psi)]_{\lambda=1} = (-1+9) \  H_{\partial/\partial t}(\psi) = 8 \ m
\end{split}
\ .
\een
On the other hand, noting that
$\delta \psi = \delta \phi + \pounds_Y \psi$ with $\delta \phi = (2g, 3A)$, and using the scaling behavior of
$\Q_{\partial/\partial t}$, we have
\ben
\begin{split}
\frac{\D}{\D \lambda} \ H_{\partial/\partial t}(\psi_\lambda) \Big|_{\lambda=1} &= \int_\infty [9 \Q_{\partial/\partial t}(\psi) - i_{\partial/\partial t} \theta(\psi; \delta \phi)] \\
& + \int_\infty [\pounds_Y \Q_{\partial/\partial t}(\psi) - i_{\partial/\partial t}(\psi; \pounds_Y \psi)] \\
& =9 \int_\infty \alpha_{\partial/\partial t} + \int_\Sigma
\pounds_Y \J_{\partial/\partial t}(\psi) - \D i_{\partial/\partial t} \theta(\psi; \pounds_Y \psi) \\
& =9 \int_\infty \alpha_{\partial/\partial t} + \int_\Sigma \omega(\psi; \pounds_Y \psi, \pounds_{\partial/\partial t} \psi) \\
& =9 \int_\infty \alpha_{\partial/\partial t} \ .
\end{split}
\een
To go to the second equality, we used the explicit expression for $\alpha_{\partial/\partial t}$ coming from eq.~\eqref{ax}, and we used Stokes
theorem to convert the second integral to that over a slice $\Sigma$. We may assume that $Y$
vanishes on the inner boundary, so there is no contribution from there. We have also used
$\D \Q_{\partial/\partial t} = \J_{\partial/\partial t}$ for the Noether current. In the third line, we have used identity~\eqref{jxom}, and in the last line we used that $\partial/\partial t$
Lie-derives $\psi$ by assumption. Combining the two formulas
for $\tfrac{\D}{\D \lambda} H_{\partial/\partial t}(\psi_\lambda)|_{\lambda=1}$, we get the formula for $m$
in the lemma. \qed

\subsection{Conservation laws and mass formulas}\label{sec:massformulas}

The Smarr relation~\eqref{smarr} derived in the previous subsection
is universal in that it holds for any stationary black hole solution
in the theory, with no extra symmetry assumptions. In this subsection we will
 show that if one makes further by-hand symmetry assumptions of the nature made in the uniqueness theorem, then one can derive {\em further non-trivial relations between the
horizon area, mass, angular momenta, electric charge etc. and the corresponding potentials
such as angular velocities of the horizon, electric potentials of the horizon, etc.} Unlike the Smarr relation, they are
not quadratic in the thermodynamic quantities.

As a difference to the previous sections, we will allow in this section
$F$'s which satisfy the Bianchi identity $\D F=0$ and field equations\footnote{Note that the field equations (although not the action) only refer to the gauge invariant field strength $F$.}, but which cannot be written globally as $F=\D A$. This means that there can be non-zero magnetic charges, which will now also enter the thermodynamic formulas. As in sec.~\ref{sec:unique}, we are assuming that the black hole spacetime is stationary with 8 additional mutually commuting rotational Killing fields, which we call $\xi_1, \dots, \xi_8$.
We assume, for definiteness that the spacetime is asymptotically Kaluza-Klein,
with 4 asymptotically large dimensions, i.e. $\M \cong \mr^{3,1} \times T^7$ in the asymptotic region. Without loss of generality we assume a labeling of the rotational Killing fields such that $\xi_8$ is
a rotation in the asymptotically large dimensions, i.e. a rotation in the $\mr^{3,1}$
near infinity, while $\xi_1, \dots, \xi_7$ are rotations in the $T^7$ near infinity. As a restriction on the class of solutions that we consider, we further assume that the black hole horizon is not rotating in the 8-direction.
In other words, if we let $K$ be the linear combination of the Killing fields pointing
along the null-generators of $\H$, then we assume
\ben
\xi_0:=K = \frac{\partial}{\partial t} + \Omega^1 \ \xi_1 + \dots + \Omega^7 \ \xi_7 \ ,
\een
so that $\Omega^8=0$.
Thus, the horizon is non-rotating in the asymptotically large dimensions,
although it can rotate in the extra-dimensions.

To make the analysis below simpler, we also impose yet further, by-hand conditions onto the
relationship between the fields $(g,F)$ and the symmetries. Our first condition is that
$\xi_8$ is hypersurface-orthogonal, or in other words
\ben\label{hsurfo}
\xi_8 \wedge \D \xi_8 = 0 \ ,
\een
where as usual, we identify vector fields with 1-forms using the metric. We can see
from the Komar expression for the angular momentum~\eqref{komar} that this
implies $J_8=0$. The second condition is
\ben\label{scal0}
i_{\xi_i} i_{\xi_j} i_{\xi_k} F = 0 \ ,
\quad
\text{for $i,j,k=1,\dots,7$.}
\een
Of course, the fields $(g,F)$ are, by assumption, also Lie-derived by any of the vector fields $\xi_0=K, \xi_1, \dots, \xi_8$. The 7 distinguished rotational
Killing fields tangent to the extra dimensions $T^7$ in the asymptotic region are
denoted by $\xi_i$, where unprimed lower case indices run from $i=1,\dots,7$. We also
adopt the convention that
lower case indices run between $i'=0,\dots,7$, and primed upper case indices from $I'=0,\dots,7,9$. As we have already described in section~\ref{sec:wp}, the horizon topology
may, in general, be either one of the possibilities given in table~\ref{table1}. But in this section, we will assume that
\ben
\B \cong S^2 \times T^7 \ ,
\een
and we also assume that $\xi_8$ is tangent to the
$S^2$-factor on the horizon\footnote{In the language of the interval structure
introduced in sec.~\ref{sec:wp}, this amounts to saying that the vectors
$v_{h-1}^i, v_{h+1}^i$ associated with the intervals adjacent to the horizon
interval are both equal to $(0,0,0,0,0,0,0,1)$.}.

As in the previous section, we get from the symmetries $X$ closed
9-forms $\alpha_{X}, X=\xi_0, \dots, \xi_8$. It turns out that there are many more closed forms, whose existence essentially follows from the sigma model formulation of this theory. We will employ these closed forms to derive our thermodynamic relations\footnote{Relations of similar nature were previously derived in~\cite{Heusler} for non-rotating black holes in Einstein-Maxwell-Dilaton theory in 4 dimensions.}. To obtain the closed forms, it turns out that
it is convenient to use a `non-compact' modification of the sigma model formulation described in sec.~\ref{sec:sigma}. The difference is that we now use the 8 Killing fields $\xi_0, \dots, \xi_7$ (spanning a timelike(!) subspace\footnote{Note that $\partial/\partial t$ itself may be spacelike (ergoregion), timelike (asymptotic region) or null (horizon, ergosurface).} in each tangent space), rather than previously $\xi_1, \dots, \xi_8$ (spanning a spacelike(!) subspace). To define the modified sigma model, we consider the 8 dimensional Gram matrix
\ben
f_{i'j'}' = g(\xi_{i'}, \xi_{j'}) \ ,
\een
which, unlike~\eqref{fdef},
is not positive definite, but has Lorentzian signature $(-1,+10)$. Similarly, we introduce the scalar potentials $\chi_{i'}', \chi_{i'j' \dots k'}'$ by formulas completely analogous
to~\eqref{chi1def} and~\eqref{chi2def}. We introduce an 8-bein by $f'_{i'j'} =
\eta_{a'b'} e_{i'}{}^{\prime a'} e_{j'}{}{}^{\prime b'}$, and write
\ben
V' = \left(
\begin{matrix}
e^{\prime a'}{}_{i'} & \det e^{\prime -1}( \chi_{i'}' + \frac{1}{720} \ \epsilon^{j_1' \dots j_8'} A_{i'j_1' j_2'}^{} \chi_{j_3' \dots j_8'}') \\
0 & \det e^{\prime -1}
\end{matrix}
\right) \ .
\een
 And we define the $\frak{e}_{8(+8)}$ valued function $v'$ by\footnote{Note that, although the potential $A$ may not be globally defined, the components $A_{I'J'K'}$ may be defined globally by the identity $\D A_{I'J'K'}
= 2\sr \ i_{\xi_{I'}} i_{\xi_{J'}} i_{\xi_{J'}} F$, because
that identity may be considered on the simply connected orbit
manifold $\hat \M$.}
\ben
v' =  e^{I'J'K'} A_{I'J'K'} - \frac{1}{360} \ e^*_{I'J'K'} \epsilon^{I'J'K'L'M'N'P'Q'R'} \chi_{L'M'N'P'Q'R'}'
\een
where $A_{I'J'K'}=-2 \sqrt{3}A_{i'j'k'}$ when $I'=i',J'=j',K'=k'$ are between $0, \dots, 7$, and
zero if $I'=9$ or $J'=9$ or $K'=9$, as well as similarly $\chi_{L'M'N'P'Q'R'}=2\sqrt{3}
\chi'_{l'm'n'p'q'r'}$ if all subscripts are between $0, \dots, 7$, and zero otherwise. We can now form the
248-dimensional matrix $\V'$ valued in the adjoint representation of the group
$E_{8(+8)}$ by
\ben
\V' = \exp({\rm ad}(v')) \ {\rm Ad}(V') \ ,
\een
[compare eq.~\eqref{Vdef}],
and we also define $N$ by
\ben
N = \V' \tau'(\V')^{-1} \ ,
\een
[compare eq.~\eqref{Mdef}].
In this equation, $\tau'$ is now the involution of $E_{8(+8)}$ defined by eq.~\eqref{taupdef}
in appendix~\ref{app:e8}. The subgroup fixed by $\tau'$ is the non-compact version $Spin^*(16)$ of $SO(16)$,
so $\V'$ may now be thought of as parameterizing the symmetric space $E_{8(+8)}/Spin^*(16)$.
The non-compact character of the subgroup $Spin^*(16)$ fixed under the involution $\tau'$
may be traced back to the timelike character of the subspace spanned by $K=\xi_0, \xi_1,
\dots, \xi_7$.

The matrix function $N$ satisfies the equations of motion of the corresponding sigma model, by complete analogy to the matrix $M$ defined before. It can easily be deduced from the equations of motion for $N$, and standard relations for Killing fields, that if we set
\ben\label{omegadef}
\omega = \star(K \wedge N^{-1} \D N) \ ,
\een
then $\omega$
is a closed 9-form, valued in the adjoint representation
of $\frak{e}_{8(+8)}$,
\ben
\D \omega = 0 \ .
\een
Hence, each of the 248 components $(\omega_{I'}{}^{J'},
\omega^{*I'J'K'}, \omega_{I'J'K'})$ in
\ben
\omega = {\rm ad}( \omega_{I'}{}^{K'} e^{I'}{}_{K'} +
\omega_{I'K'L'} e^{I'K'L'} + \omega^{*I'K'L'} e^*_{I'K'L'} )
\een
is a scalar valued, closed 9-form on $\M$.
(Here, the star $*$ on the symbols is part of the name and does not
mean any kind of conjugation or dual.)  Their
concrete expression is very lengthy and is given in \eqref{om1},~\eqref{om2},~\eqref{om3} of appendix~\ref{app:J}.
The ``conservation laws'' $\D \omega=0$ are, in essence, the Noether currents of the hidden symmetry $E_{8(+8)}$ of the field equations, which is made manifest in the sigma model formulation.

We now integrate
$0 = \D \omega$ over a 10-dimensional surface $\Sigma$ going from the bifurcation surface
$\B$ of the horizon to spatial infinity. Using Stokes theorem, we then get
the 80 relations
\ben\label{j1}
\int_\B \omega_{I'}{}^{J'} = \int_\infty \omega_{I'}{}^{J'} \ ,
\een
the 84 relations
\ben\label{j2}
\int_\B \omega_{I'K'L'} = \int_\infty \omega_{I' K' L'} \ ,
\een
and the 84 relations
\ben\label{j3}
\int_{\B} \omega^{*I'K'L'} =
\int_\infty \omega^{*I'K'L'}  \ .
\een
The key step is now to relate the quantities on both sides with the thermodynamic quantities characterizing the black hole. These are as follows: For the mass $m$ and angular momenta $J_i$ we may use the Komar expressions~\eqref{komar}.  The electric charge $Q[C_7]$
associated with a 7-cycle was defined above in eq.~\eqref{Qdef}. The 7-cycles lying
in $\H \cong \mr \times (S^2 \times T^7)$ carrying a non-zero electric charge are $C_7 \cong S^2 \times C_5$, where $C_5 \subset T^7$
is an embedded 5-torus. The position of $C_5$ is characterized by 5 generators $\xi_{i_1},
\dots, \xi_{i_5}$, so the 7-cycle $C_7$ is characterized by $[i_1 \dots i_5]$. We use the shorthand
\ben
Q_{klmnr} = (2\pi)^2 \int_{[klmnr]} 4 \ q \ .
\een
We also define $Q^{ij}=\frac{1}{5!} \epsilon^{ijklmnr} Q_{klmnr}$, which obviously
contains the same information.
Using the relation $\pounds_X = i_X \D + \D i_X$, the Bianchi
identity $\D F = 0$, and the fact that $\xi_i, \xi_j, K$
Lie-derive $F$, it follows that $\D(i_{\xi_i} i_{\xi_j} i_K F)=0$ on $\H$, hence there is a scalar function\footnote{Even though
$\M$ is not simply connected, the forms under consideration
may be viewed as forms on the simply connected orbit space $\hat M$,
so there is no difference between closed and exact invariant 1-forms.} such that
\ben\label{phidef}
\D \Phi_{ij}
= - i_{\xi_i} i_{\xi_j} i_K F \ .
\een
Furthermore, we already showed in sec.~\ref{sec:firstlaw} that
$i_K F=0$ on $\H$, so $\Phi_{ij}$
is constant on $\H$. The constant is fixed by demanding that $\Phi_{ij}$ vanishes at infinity. Finally, we consider the non-trivial 4-cycles in
$\H$ of the form $C_4 \cong S^2 \times C_2$, where $C_2 \subset T^7$ is an embedded 2-torus.
Its position is characterized by 2 generators $\xi_{i_1}, \xi_{i_2}$, so the corresponding
4-cycle is characterized by $[i_1 i_2]$. The associated magnetic
charges are then defined as
\ben
P_{ij} = (2\pi)^5 \int_{[ij]}  F \ .
\een
If $F$ happens to be equal $\D A$ for a globally defined 3-form $A$, then of course the magnetic charges are zero. Now take 5 Killing fields $\xi_i, \dots, \xi_k$, and form the 1-form $i_K i_{\xi_k} \dots i_{\xi_r} q$. Because
$q$ is closed, so is this 1-form, and we get, by the same argument as above, a scalar function which we may call
\ben\label{psidef}
\D \Psi^{ij} = - \frac{1}{5!} \epsilon^{ijklmnr} \ i_K i_{\xi_r} \dots i_{\xi_k} q \ ,
\een
where we mean the 7 dimensional totally antisymmetric tensor (recall that lower case Roman indices
$i=1, \dots, 7$ in this section).
Actually, $\Psi^{ij}$, the magnetic potentials, are again constant on $\H$; one proves this by the exactly the same argument as just given
for the electric potentials.
Thus, in summary,
our thermodynamical quantities are the mass and angular momenta $m,J_i$, the angular
velocities of the horizon, $\Omega^i$, the electric/magnetic potentials $\Phi_{ij}, \Psi^{ij}$, the electric/magnetic charges $Q^{ij}, P_{ij}$, the horizon area $\A$, and the
surface gravity $\kappa$. We now give relations between them following from
eqs.~\eqref{j1},~\eqref{j2},~\eqref{j3}; the lengthy calculations are outlined in appendix~\ref{app:J}.

Evaluating eq.~\eqref{j2} with the choice $I'=0, J'=j, K'=k$ gives
\ben
\begin{split}
&-d_{jl} d_{km} \ Q^{lm} =-2 \ \Phi_{jk} \kappa \A + 2 \ \Psi^{mn} \Phi_{mn} P_{jk} \\
&+ 4 \ \Phi_{l[k} \Phi_{j]m} Q^{lm} - 8 \ \Phi_{l[j} \Psi^{ml} P_{k]m} +
4 \ \Phi_{jk} \Psi^{lm} P_{lm}   \  \ .
\end{split}
\een
Here, we remind the reader of our convention that lower case Roman indices
go from $i=1, \dots, 7$, and we have also introduced the 7 by 7 matrix $d^{ij}$
by
\ben
d^{ij} = \delta^{ij} - \Omega^i \Omega^j \ ,
\een
with $d_{ij}$ denoting its inverse. Next, evaluating eq.~\eqref{j3} with $I'=9, J'=j, K'=k$
gives
\ben
\begin{split}
& -d^{jl} d^{km} \ P_{lm} = -2 \ \Psi^{jk} \kappa \A + 2 \ \Psi^{mn} \Phi_{mn} Q^{jk} \\
&+ 4 \ \Psi^{l[k} \Psi^{j]m} P_{lm}
- 8 \ \Psi^{l[j} \Phi_{ml} Q^{k]m}
+ 4 \ \Psi^{jk} \Phi_{lm} Q^{lm}    \ \ .
\end{split}
\een
Taking $I'=0=J'$ in eq.~\eqref{j1} gives
\ben
9 \cdot 2  \kappa \A = 8 \ m - 9 \ \Omega^i J_i + 12 \  \Phi_{ij} Q^{ij} + 6 \ \Psi^{ij} P_{ij}  \ ,
\een
which is the Smarr relation~\eqref{smarr}, already derived earlier in the absence of magnetic charges. Taking $I'=0, J'=j$ in eq.~\eqref{j1} gives
\ben
-\frac{1}{4} \ \epsilon^{jikmnpq} \ \Phi_{ik} \Phi_{mn} P_{pq} = -9 \ \delta^{jk} J_k + 8 \ m \Omega^j \ .
\een
Taking $I'=i$ and $J'=9$ in eq.~\eqref{j1} gives
\ben
0 = \frac{1}{6} \ \epsilon_{ijkpqmn} \Psi^{jk} \Psi^{pq} Q^{mn} 
+ 4 \ \Phi_{ij} \Psi^{jk} J_k \ .
\een
Taking $I'=0, J'=9$ in eq.~\eqref{j1} gives
\ben
\Phi_{mn}\Psi^{mn} \ \kappa\A = 4 \ \Psi^{pq} \Phi_{pl} \Phi_{mq} Q^{lm} + 4 \ \Phi_{pq} \Psi^{pl} \Psi^{mq} P_{lm} \ .
\een
Finally, taking $I'=i, J'=j$ in eq.~\eqref{j1} gives
\ben
4\pi \ d^{jm} \tau_{im} + \Omega^j J_i = -4 \ \Psi^{jm} P_{im} + \frac{2}{3} \ \delta^j{}_i \Psi^{mn} P_{mn}
 - 4 \ \Phi_{im} Q^{jm} + \frac{2}{3} \ \delta^j{}_i \Phi_{mn} Q^{mn} \ .
\een
Here, the 7 by 7 constant matrix $\tau_{mn}$ is defined by the relation
\ben
f_{ij} = \delta_{ij} - \frac{1}{(2\pi)^7} \ \frac{\tau_{ij}}{R} + O(R^{-2}) \ ,
\een
where $R=\sqrt{x_1^2 + x_2^2 + x_3^2}$ is the standard radial coordinate in the large
dimensions ($\mr^{3,1}$) relative to an asymptotically Cartesian coordinate system. It is physically
interpreted as the ``tension tensor'' of the 7 asymptotically small dimensions\footnote{These are the ``thermodynamic potentials'' which would be conjugate to the deformations of the asymptotic
metric on $T^7$ if we would loosen our boundary conditions
to allow such. They would give rise to a corresponding
term in the first law, see e.g.~\cite{traschen} for an example.} ($T^7$).

We emphasize that all these thermodynamic formulas have not been obtained from a particular explicit solution, but from the general structure of the equations of motion (hidden symmetries) for the class of solutions which have the indicated symmetries. We leave for the future a more detailed analysis of these relations.

\section{Conclusions}

In this paper, we have considered stationary solutions in 11-dimensional supergravity theory.
We first derived the first law of black hole mechanics, valid for arbitrary horizon topologies, and arbitrary stationary black holes, without additional symmetry assumptions. We then specialized to stationary (asymptotically Kaluza-Klein) solutions whose isometry group is (or contains) $\mr \times U(1)^8$. In this case, we were able to associate with each such solution a collection of moduli and generalized winding numbers which encode the topology of the solution and the action of the isometries. Furthermore, for each given set of moduli, generalized winding numbers, angular momenta, and electric type charges,
we proved a black hole uniqueness theorem.

The proof of this theorem makes use of the known sigma-model formulation of the field equations of 11-dimensional supergravity when it is ``dimensionally reduced''. The sigma model formulation also has another application explored in this paper, namely it gives an interesting set of relations between the electric/magnetic charges and potentials, angular momenta and velocities, mass, area and surface gravity. These, rather non-trivial, relations generalize the well-known Smarr-type formulas. We believe that there is a relation between these
formulas and the formulas given in~\cite{nicolai} for the solutions corresponding to ``nilpotent orbits''. However, this remains to be worked out.

\vspace{1cm}

{\bf Acknowledgements:} I would like to thank R.M.~Wald for discussions about the
first law in the presence of magnetic charges, and H.~Nicolai for discussions about
coset formulations of supergravity and nilpotent orbits. The author is supported by
ERC starting grant no. QC \& C 259562.

\appendix

\section{Calculation of the Noether charge and constraints}
\label{app:nc}
In this appendix, we derive the expressions for the Noether charge and constraints quoted in sec.~\ref{sec:covphase}. The Lagrange 11-form is
given in components by
\ben
\begin{split}
& L_{a_1 \dots a_{11}} \\
=& \left( R - \frac{1}{12}\ F_{bcde} F^{bcde} + \frac{1}{2,592}\ \epsilon^{b_1 \dots b_{11}} F_{b_1b_2b_3b_4} F_{b_5 b_5 b_7 b_8} A_{b_9 b_{10} b_{11}}
\right)  \ \epsilon_{a_1 \dots a_{11}} \ .
\end{split}
\een
where the epsilon tensor is the natural volume element defined from the metric relative to a given orientation of $\M$. The 10-form $\theta$ is given by
\bena
&&\theta_{a_1 \dots a_{10}} = \epsilon_{da_1 \dots a_{10}} v^d \ , \\
&& v_d = g^{bc}(\nabla_c \delta g_{bd} - \nabla_d \delta g_{bc}) -
\frac{1}{3} \ F_d{}^{bce} \delta A_{bce} + \frac{1}{324} \ \epsilon_d{}^{b_1 \dots b_{10}} \delta A_{b_1 b_2 b_3} F_{b_4 b_5 b_6 b_7} A_{b_8 b_9 b_{10}} \ .
\eena
From~\eqref{jxdef}, one obtains the following expression for the components of the Noether current 10-form (here we drop the subscript ``$X$'' on $\J_X$):
\ben
\begin{split}
\J_{a_1 \dots a_{10}} &= -2 \ \epsilon_{da_1 \dots a_{10}} \nabla_e( \nabla^{[d} X^{e]}) \\
&+ \epsilon_{da_1 \dots a_{10}} \bigg(
2 \ G_e{}^d X^e + \frac{1}{12} \ F_{bcfg} F^{bcfg} X^d
-\frac{1}{324} \ \epsilon^{b_0 b_1 \dots b_{10}} F_{b_0 b_1 b_2 b_3} F_{b_4 b_5 b_6 b_7} A_{b_8 b_9 b_{10}} X^d
\bigg)\\
&+ \epsilon_{da_1 \dots a_{10}} \bigg(
-\frac{2}{3} \ F^{dbcf} (X^e \nabla_e A_{bcf} + 3\ A_{bce} \nabla_f X^e)\\
&\hspace{3cm} + \frac{1}{324} \ \epsilon^{db_1 \dots b_{10}} (X^e \nabla_e A_{b_1b_2b_3}
+ 3 \ A_{eb_2 b_3} \nabla_{b_1} X^e) F_{b_4 b_5 b_6 b_7} A_{b_8 b_9 b_{10}}
\bigg) \ .
\end{split}
\een
We rewrite the term $\nabla_e A_{b_1b_2b_3}$ in the last line as
\ben
\nabla_e A_{abc} = F_{eabc} + \nabla_a A_{ebc} + \nabla_b A_{eca} +
\nabla_c A_{eab} \ .
\een
Differentiating by parts and rearranging, we obtain:
\ben
\begin{split}
\J_{a_1 \dots a_{10}}
&=2 \ \epsilon_{da_1 \dots a_{10}} (
t_e{}^d X^e - j^{dbc} A_{ebc} X^e ) \\
&+\epsilon_{da_1 \dots a_{10}} \nabla_e \bigg(
-2 \  \nabla^{[d} X^{e]} -2 \ F^{debc} A_{fbc} X^f
+ \frac{1}{108} \ \epsilon^{def_1 \dots f_9} A_{cf_1 f_2} F_{f_3 f_4 f_5 f_6}
A_{f_7 f_8 f_9} X^c
\bigg) \ ,
\end{split}
\een
where we have defined
\ben\label{tjdef}
\begin{split}
&t_{ab} = G_{ab} - T_{ab} \\
&j^{bcd} = \nabla_a F^{abcd} - \frac{1}{576} \ \epsilon^{bcdf_1 \dots f_8}
F_{f_1 f_2 f_3 f_4} F_{f_5 f_6 f_7 f_8} \ ,
\end{split}
\een
and where
\ben\label{tabdef}
T_{ab} = \frac{1}{3} \ F_{acde} F_b{}^{cde} - \frac{1}{24} \
g_{ab} F_{cdef} F^{cdef} = \frac{1}{6} \ F_{acde} F_b{}^{cde} +
\frac{1}{720} \ (\star F)_{acdefgh} (\star F)_b{}^{cdefgh}
\een
is the ``electromagnetic'' stress tensor.
$t_{ab}$ is interpreted as the non-gravitational stress energy tensor, because
it is just the difference between the Einstein tensor and the electromagnetic stress tensor. $j^{abc}$ is interpreted as the non-electromagnetic current.
Of course, $t_{ab} = 0 = j^{abc}$ when the equations of motion hold. Therefore, we can read off the constraints $\C_X$ and Noether charge $\Q_X$ from the expression for the Noether current as (we drop the subscript `$X$'):
\ben
\begin{split}
&\C_{a_1 \dots a_{10}} = \epsilon_{da_1 \dots a_{10}} \bigg(
2\ t_e{}^d X^e - 2 \ j^{dbc} A_{ebc} X^e \bigg) \\
& \Q_{a_2 \dots a_{10}} = \epsilon_{dea_2 \dots a_{10}} \bigg(
-\nabla^{[d} X^{e]} - \ F^{debc} A_{fbc} X^f
+ \frac{1}{216} \ \epsilon^{def_1 \dots f_9} A_{cf_1 f_2} F_{f_3 f_4 f_5 f_6}
A_{f_7 f_8 f_9} X^c
\bigg) \ .
\end{split}
\een
These expressions can be conveniently rewritten in differential forms notation if we define the components of the 1-form $f_X$ by
\ben\label{fxdef}
f_a = t_{ab} X^b \ ,
\een
(we drop the subscript `$X$') as well as, as usual,
\ben
(\star j)_{a_1 \dots a_8} = \frac{1}{3!} \epsilon_{bcda_1 \dots a_8} \  j^{bcd} \ .
\een
Then we get
\ben\label{noesugra1}
\begin{split}
\Q_X &= -\star \D X - 4 \ i_X A \wedge q + \frac{4}{3} \ i_X A \wedge A \wedge F \ , \\
\C_X &= 2 \ \star f_X + 4 \ i_X A \wedge \star j \ .
\end{split}
\een
where $\star j = \D q$. These are the formulas claimed in the main text.
As in the main text, we
use the standard operators $(\D \alpha)_{a_1...a_p} = p
\nabla_{[a_1}\alpha_{a_2...a_p]}$ (the exterior derivative),
$(\alpha \wedge \beta)_{a_1 ... a_p b_1 ... b_q} =
\frac{(p+q)!}{p!q!} \alpha_{[a_1 ... a_p} \beta_{b_1...b_q]}$ (the wedge product), $(\star \alpha)_{a_1...a_p} = \frac{1}{(n-p)!} \epsilon_{b_1...b_{n-p}a_1...a_p} \alpha^{b_1...b_{n-p}}$ (the Hodge
dual), and $(i_X \alpha)_{a_1...a_p} = X^b \alpha_{ba_1...a_p}$ (the
interior derivative).

\section{Basic facts and definition of $E_{8(+8)}$}\label{app:e8}

The exceptional, real, Lie-algebra $\frak{e}_{8(+8)}$ is a particular real form of the
complex exceptional Lie-algebra $\frak{e}_8$. This semi-simple
Lie-algebra can be characterized by a Cartan-matrix with corresponding generators and relations, but for our purposes, another set of generators is more suitable. These
are often referred to as `Freudenthal's realization'~\cite{freudenthal}, and
are denoted by $e^I{}_J, e_{IJK}, e^{*IJK}$, $I,J,K=1,\dots,9$. They are subject to the following relations. Both $e_{IJK}, e^{*IJK}$ are totally antisymmetric in the indices, and
$e^J{}_J=0$. The 80 basis elements $e^I{}_J$ generate the Lie-algebra
$\frak{sl}(9)$,
\ben
[e^I{}_J, e^K{}_L] = \delta^I{}_L e^K{}_J - \delta^K{}_J e^I{}_L \ .
\een
The following relations manifest how the adjoint representation $\frak{e}_{8(+8)}$ splits~\eqref{decomp}
under the restriction to $\frak{sl}(9)$,
\ben
\begin{split}
&[e_{IJK}^*, e^L{}_M] = \delta^L{}_{I} e_{MJK}^* + \delta^L{}_{J} e_{IMK}^* + \delta^L{}_{K} e_{IJM}^*\\
&[e^{IJK}, e^L{}_M] = -\delta_M{}^{I} e^{LJK} -\delta_M{}^{J} e^{ILK} -\delta_M{}^{K} e^{IJL} \ .
\end{split}
\een
 The remaining brackets are
\ben
\begin{split}
&[e^{IJK}, e^{LMN}] = \frac{1}{36\sqrt{3}} \ \epsilon^{IJKLMNPQR} e_{PQR}^*  \\
&[e_{IJK}^*, e_{LMN}^*] = \frac{1}{36\sqrt{3}} \ \epsilon_{IJKLMNPQR} e^{PQR}  \\
&[e_{IJK}^*, e^{LMN}] = -\frac{1}{6} \ \delta^L{}_{[I} \delta^M{}_{J} e^N{}_{K]} \ .
\end{split}
\een
The {\em real} span of these generators is by definition  the Lie algebra $\frak{e}_{8(+8)}$, whereas the complex span is $\frak{e}_8$. Its dimension is 80+84+84=248.

The Lie-algebra $\frak{e}_{8(+8)}$ has several involutions, i.e. Lie-algebra automorphisms $\tau$ (meaning
 $\tau([X,Y]) = [\tau(X), \tau(Y)]$) such that $\tau^2 = id$. In this paper, we consider two of them. The first one is defined by
\ben\label{taudef}
\tau(e^I{}_J) = -\delta^{IK} \delta_{JL} e^L{}_K \ , \quad
\tau(e^{*IJK}) = \delta^{IL} \delta^{JL} \delta^{KN} e_{LMN} \ , \quad
\tau(e_{IJK}) = \delta_{IL} \delta_{JL} \delta_{KN} e^{*LMN} \ ,
\een
where $\delta_{IJ} = diag(1,1,1,1,1,1,1,1,1)$ is the 9-dimensional Euclidean metric. The second one is
\ben\label{taupdef}
\tau'(e^I{}_J) = -\eta^{IK} \eta_{JL} e^L{}_K \ , \quad
\tau'(e^{*IJK}) = \eta^{IL} \eta^{JL} \eta^{KN} e_{LMN} \ , \quad
\tau'(e_{IJK}) = \eta_{IL} \eta_{JL} \eta_{KN} e^{*LMN} \ ,
\een
where $\eta_{IJ}= diag(-1,1,1,1,1,1,1,1,-1)$ is a 9-dimensional flat pseudo-Riemannian metric. The elements left invariant by an automorphism automatically form a subalgebra. In the case of $\tau$, this can be seen to be $\frak{so}(16)$, whereas in the case of $\tau'$, this can be seen to be\footnote{It is fairly obvious that this Lie-algebra, $\frak{g}^{\tau'}$, must be
a real form of $\mc \otimes \frak{so}(16)$. To see that it must in fact be $\frak{so}^*(16)$,
one can verify that the restriction of the Cartan-Killing form of $\frak{e}_{8(+8)}$
 to $\frak{g}^{\tau'}$ has signature $(-64,+56)$. By identifying the
 the generators corresponding to the $64$ negative signs, one sees that these
 correspond to the Lie-algebra $\mathfrak{u}(8)$, which is a maximal compact
 sub-algebra of $\frak{so}^*(16)$.} $\frak{so}^*(16)$. The connected Lie-group
corresponding to $\frak{e}_{8(+8)}$ is denoted by $E_{8(+8)}$. The corresponding group
automorphisms are denoted, by abuse of notation, by the same symbols $\tau, \tau'$.
The triples $(E_{8(+8)}, SO(16), \tau)$ resp. $(E_{8(+8)}, Spin^*(16), \tau')$ form symmetric spaces. The subgroups clearly have dimension 120, so the dimension of the coset
spaces $E_{8(+8)}/SO(16)$ and $E_{8(+8)}/Spin^*(16)$ is hence 128.

\section{Asymptotic conditions}\label{app:ac}

Asymptotically KK-boundary conditions are in more detail as follows:
We assume that a subset  of $\M$ is diffeomorphic to the cartesian product of
$\mr^s$ with a ball removed---corresponding to the asymptotic region of the large spatial
dimensions---and $\mr \times \T^{D-s-1}$---corresponding to the time-direction and small dimensions.
We will refer to this region as the asymptotic region and call it $\M_{\rm asymptotic}$.
The metric is required to behave in this region like
\ben\label{standard}
g = -\D t^2 + \sum_{i=1}^s \D x_i^2 + \sum_{i=1}^{10-s} \D \phi_i^2 + O(R^{-s+2}) \, ,
\een
where $O(R^{-\alpha})$ stands for metric components that drop off faster than $R^{-\alpha}$ in the radial
coordinate $R = \sqrt{x_1^2+...+x_s^2}$, with $k$-th derivatives in the coordinates
$x_1, \dots, x_s$ dropping off at least as fast as
$R^{-\alpha-k}$. These terms are also required to be independent of the coordinate
$t$, which together with $x_i$ forms the standard  cartesian coordinates on $\mr^{s,1}$.
The remaining coordinates $\phi_i$ are $2\pi$-periodic and parameterize the torus
$\T^{D-s-1}$. The timelike Killing field is assumed to be equal to $\partial/\partial t$ in
$\M_{\rm asymptotic}$.
We also require that the 3-form field has asymptotic behavior
\ben
\begin{split}
A =& \sum_{i,j,k=1}^{10-s} O(1) \D \phi^i \wedge \D \phi^j \wedge \D \phi^k +
\sum_{i,j=1}^{10-s} \sum_{\mu=0}^s O(R^{-s+2}) \D x^\mu \wedge \D \phi^i \wedge \D \phi^j + \\
&\sum_{i=1}^{10-s} \sum_{\mu=0}^s O(R^{-s+2}) \D x^\mu \wedge \D x^\nu \wedge \D \phi^j
+\sum_{\mu,\nu,\sigma=0}^s O(R^{-s+2}) \D x^\mu \wedge \D x^\nu \wedge \D x^\sigma \ ,
\end{split}
\een
with all components independent of $t$. In sections~\ref{sec:mform},~\ref{sec:unique}
we make the more restrictive assumption that the first term
on the right side is $O(R^{-s+1})$. This is done mainly for simplicity. Otherwise, the asymptotic values for $A_{ijk}$ at infinity appear as additional parameters in the thermodynamic relations.
We call spacetimes satisfying these properties
``asymptotically Kaluza-Klein spacetimes''\footnote{For the axisymmetric spacetimes considered in
this paper, one can derive more precise asymptotic expansions, as explained in~\cite{HollandsYazadjiev08b} for the example of the vacuum field equations.}.

\medskip

Unfortunately, in order to make many of the
arguments in the body of the paper in a consistent way, one has to make certain further technical
assumptions about the global nature of $(\M,g)$ and the action of the symmetries.
Our assumptions are in parallel to those made by Chrusciel and Costa
in their study~\cite{CC08} of 4-dimensional stationary black holes. The
requirements are (a) that $\M$ contains an acausal, spacelike, connected
hypersurface $\Sigma$ asymptotic to the $t = 0$ surface in the asymptotic region,
whose closure has as its boundary $\partial \Sigma = \B$ a cross section
of the horizon. We always assume $\B$ to be compact and connected.
(b) We assume that the orbits of $\partial/\partial t$ are complete. (c) We assume that
the horizon is non-degenerate. (d) We assume that $\M$
is globally hyperbolic. In order to use the rigidity theorem in sec.~\ref{sec:firstlaw}, and to prove the orbit space theorem in sec.~\ref{sec:wp}, it is necessary to assume (e) that the spacetime, the metric, and
the group action are analytic, rather than only smooth.

\section{Formulas for $\omega$}\label{app:J}

In this section, we give the concrete expression for the closed 9-forms $\omega_{I'}{}^{J'},
\omega^{*I'J'K'}, \omega_{I'J'K'}$ defined by eq.~\eqref{omegadef}. As in sec.~\ref{sec:mform}, we use the index conventions that lower case primed indices run between $i'=0,1,\dots,7$,
lower case unprimed indices run between $i=1,\dots,7$, and upper case primed indices
run between $I'=0,1,\dots,7,9$. And again, we assume a labeling of the Killing fields
such that $\xi_0=K$ is the Killing field tangent to the null generators of $\H$, such that $\xi_8$ is tangent to the $S^2$ factor of $\H \cong \mr \times S^2 \times T^7$, and such
that $\xi_1, \dots, \xi_7$ are tangent to the extra dimensions $\cong T^7$ in the asymptotic
region. We define
\ben
\varphi^{I'J'K'} = -\frac{\sqrt{3}}{360} \epsilon^{I'J'K'l'm'n'p'q'r'} \chi'_{l'm'n'p'q'r'} \ ,
\een
and as before $A_{I'J'K'} = -2 \sqrt{3} A_{i'j'k'}$ when all indices are between $0,\dots,7$
and 0 if one index is $=9$. Let us then define the following 9-forms  $k_{I'}{}^{J'}$,
\ben
\begin{split}
k_{i'}{}^{j'} &=
f^{j'k'} \star\!(K\wedge \D f_{i'k'}) \\
k_{9 \,}{}^9 &=- f^{m'n'} \star \!(K \wedge \D f_{m'n'}) \\
k_{i'}{}^9 &= 2 \ f_{m'(k'} U_{i')}  \ \star\!(K \wedge \D f^{k'm'})\\
k_{9 \,}{}^{j'} &= 0 \ ,
\end{split}
\een
where $U_{i'} = \chi_{i'}' + \frac{1}{720} \ \epsilon^{j_1' \dots j_8'} A_{i'j_1' j_2'}^{} \chi_{j_3' \dots j_8'}'$. Furthermore, the 9-forms $k_{I'J'K'}$ are defined by
\ben
\begin{split}
k_{i'j'k'} =& -2\sqrt{3} \ i_K \star \!(F \wedge \xi_{k'} \wedge \xi_{j'} \wedge \xi_{i'})\\
&+ \frac{\sqrt{3}}{360}\frac{1}{\det f'} \ U_{i'} f_{j'm'} f_{k'n'}
\epsilon^{m'n'p' \dots q'} i_K(F \wedge \xi_{q'} \wedge \dots \wedge \xi_{p'})\\
&+ \frac{\sqrt{3}}{180} \frac{1}{\det f'}\ U_{[j'} f_{k']m'} f_{i'n'}
\epsilon^{m'n'p' \dots q'} i_K(F \wedge \xi_{q'} \wedge \dots \wedge \xi_{p'})\\
k_{9j'k'} =& +\frac{\sqrt{3}}{360} \ \frac{1}{\det f^{\prime}} \ f_{j'm'} f_{k'n'}
\epsilon^{m'n'p' \dots q'} i_K(F \wedge \xi_{q'} \wedge \dots \wedge \xi_{p'})
\end{split}
\een
and the 9-forms $k^{*I'J'K'}$ are defined by
\ben
\begin{split}
k^{*i'j'k'} =& -2\sqrt{3} \ f^{i'p'} f^{j'q'} f^{k'r'} i_K \star \!(F \wedge \xi_{r'} \wedge
\xi_{q'} \wedge \xi_{p'})\\
k^{*9j'k'} =& \frac{\sqrt{3}}{360} \ \epsilon^{j'k'p'\dots q'} i_K(F \wedge \xi_{q'}
\wedge \dots \wedge \xi_{p'})\\
&+2\sqrt{3} \ U_{m'} f^{m'n'} f^{j'p'} f^{k'q'}i_K(F \wedge \xi_{q'}
\wedge \xi_{p'} \wedge \xi_{n'})
 \ .
\end{split}
\een
In these formulas, we have, as usual, identified the vector fields $\xi_{i'}$ with
1-forms using the metric.
The formulas for the 9-forms  $\omega_{I'}{}^{J'},
\omega^{*I'J'K'}, \omega_{I'J'K'}$ are then:
\ben
\begin{split}
\label{om1}
&\omega_{I'}{}^{J'}  = \\
&+k_{I'}{}^{J'} - \frac{1}{12} \ (A_{I'P'Q'} \varphi^{L'P'Q'}\delta_{M'}{}^{J'} + A_{M'P'Q'}\varphi^{J'P'Q'} \delta_{I'}{}^{L'}\\
&+4 \ A_{I'M'P'} \varphi^{J'L'P'} - \frac{2}{3} \ A_{M'P'Q'} \varphi^{L'P'Q'} \delta_{I'}{}^{J'} - \frac{2}{3} \ A_{I'P'Q'} \varphi^{J'P'Q'} \delta_{M'}{}^{L'}) \ k_{L'}{}^{M'}\\
& +\frac{1}{432\sqrt{3}} (\epsilon^{J'S'N'P'Q'R'L'U'V'} A_{I'S'N'} A_{P'Q'R'} A_{M'U'V'}
+ \epsilon_{I'S'N'P'Q'R'L'U'V'} \varphi^{J'S'N'} \varphi^{P'Q'R'} \varphi^{L'U'V'}) \ k_{L'}{}^{M'} \\
& -\frac{1}{144} \ \varphi^{J'P'Q'} \varphi^{L'U'V'}(A_{I'P'Q'}A_{M'U'V'} - 4 \ A_{I'P'V'}
A_{M'U'Q'}) k_{L'}{}^{M'}\\
& +\frac{1}{6} \ (\varphi^{J'M'N'} \delta_{I'}{}^{L'} - \frac{1}{9} \ \delta_{I'}{}^{J'} \varphi^{L'M'N'})k_{L'M'N'} -
\frac{1}{432 \sqrt{3}} \ \epsilon^{J'V'W'P'Q'R'L'M'N'} A_{I'V'W'} A_{P'Q'R'} k_{L'M'N'} \\
&+\frac{1}{36} \ (\frac{1}{6} \ A_{I'P'Q'} \varphi^{J'P'Q'} \varphi^{L'M'N'} - A_{I'P'Q'} \varphi^{J'P'L'} \varphi^{M'N'Q'} + A_{P'Q'R'} \varphi^{P'Q'L'} \varphi^{M'J'R'} \delta_{I'}{}^{N'}) k_{L'M'N'} \\
&+\frac{1}{432 \sqrt{3}} \ A_{S'T'U'} \varphi^{J'V'W'} k_{L'M'N'} \epsilon^{L'M'N'P'Q'R'S'T'U'V'}(\frac{1}{3} \ A_{I'Q'R'} A_{P'V'W'} + \frac{2}{3} \
A_{I'V'P'} A_{Q'R'W'})\\
&-\frac{1}{6} \ (A_{I'M'N'} \delta_{L'}{}^{J'} - \frac{1}{9} \ \delta_{I'}{}^{J'} A_{L'M'N'})
k^{*L'M'N'} \\
&+\frac{1}{432\sqrt{3}} \ \epsilon_{I'V'W'P'Q'R'L'M'N'} \varphi^{J'V'W'} \varphi^{P'Q'R'} k^{*L'M'N'} \\
&-\frac{1}{36} \ (\frac{1}{6} \ \varphi^{I'P'Q'} A_{I'P'Q'} A_{L'M'N'} - \varphi^{J'P'Q'} A_{I'P'L'} A_{M'N'Q'}) k^{*L'M'N'} \\
&+\frac{1}{432\sqrt{3}} \ \epsilon^{J'V'W'P'Q'R'S'T'U'} A_{I'V'W'} A_{P'M'N'} A_{L'Q'R'} A_{S'T'U'} k^{*L'M'N'} \ .
\end{split}
\een
Furthermore,
\ben
\label{om2}
\begin{split}
& \omega_{I'J'K'} =\\
&-3 \ (A_{M'[J'K'} k_{I']}{}^{M'} - \frac{1}{9} \ A_{I'J'K'} k_{M'}{}^{M'})\\
&+\frac{1}{24\sr} \ \epsilon_{I'J'K'P'Q'R'M'U'V'} \varphi^{L'U'V'}\varphi^{P'Q'R'}k_{L'}{}^{M'}\\
&+\frac{1}{6} \ (-\frac{1}{2} \ \varphi^{L'P'Q'} A_{M'P'Q'} A_{I'J'K'} k_{L'}{}^{M'}
+ 3 \ \varphi^{L'P'Q'} A_{M'P'[I'} A_{J'K']Q'} k_{L'}{}^{M'})\\
&-\frac{1}{24\sr} \ \epsilon^{L'V'W'P'Q'R'S'T'U'} A_{M'V'W'} A_{P'[J'K'} A_{I']Q'R'} A_{S'T'U'} k_{L'}{}^{M'} + k_{I'J'K'}\\
&+\frac{1}{2} \ (\frac{1}{2} \ A_{P'Q'[I'} \varphi^{L'P'Q'} \delta_{J'}{}^{M'} \delta_{K']}{}^{N'} k_{L'M'N'} - A_{P'[I'J'} \varphi^{L'M'P'} \delta_{K']}{}^{N'} k_{L'M'N'} \\
&+\frac{1}{9} \ A_{I'J'K'} \varphi^{L'M'N'} k_{L'M'N'})\\
&+\frac{1}{72\sr} \ \epsilon^{L'M'N'P'Q'R'S'T'U'} A_{P'[J'K'} A_{I']Q'R'} A_{S'T'U'} k_{L'M'N'} \\
&+\frac{1}{36\sr} \ \epsilon_{I'K'J'P'Q'R'L'M'N'} \varphi^{P'Q'R'} k^{*L'M'N'} \\
&+\frac{1}{4} \ (A_{L'[J'K'} A_{I']M'N'} - \frac{1}{9} \ A_{I'J'K'} A_{L'M'N'}) k^{*L'M'N'} \ ,
\end{split}
\een
and finally,
\ben
\label{om3}
\begin{split}
& \omega^{*I'J'K'} =\\
&-3 \ (\varphi^{L'[J'K'} k_{L'}{}^{I']} - \frac{1}{9} \ \varphi^{I'J'K'} k_{L'}{}^{L'})
-\frac{1}{24\sqrt{3}} \epsilon^{I'J'K'P'Q'R'L'U'V'} A_{M'U'V'} A_{P'Q'R'} k_{L'}{}^{M'}\\
&-\frac{1}{6} \ (-\frac{1}{2} \ A_{M'P'Q'} \varphi^{L'P'Q'} \varphi^{I'J'K'} k_{L'}{}^{M'} + 3 \ A_{M'P'Q'} \varphi^{L'P'[I'}
\varphi^{J'K']Q'} k_{L'}{}^{M'} \\
&- 3 \ A_{P'Q'R'} \varphi^{P'Q'[I'}
\delta_{M'}{}^{J'} \varphi^{K']L'R'} k_{L'}{}^{M'})\\
&-\frac{1}{24\sqrt{3}} \ \epsilon^{I'J'K'P'Q'R'S'T'U'}
(\frac{1}{3} \ A_{M'Q'R'} A_{P'V'W'} + \frac{2}{3} \ A_{M'V'P'} A_{Q'R'W'})A_{S'T'U'} \varphi^{L'V'W'} k_{L'}{}^{M'} \\
&+\frac{1}{36\sqrt{3}} \ \epsilon^{I'J'K'P'Q'R'L'M'N'}
A_{P'Q'R'} k_{L'M'N'} \\
&+\frac{1}{4} \ (\varphi^{L'[J'K'} \varphi^{I']M'N'} - \frac{1}{9} \
\varphi^{I'J'K'} \varphi^{L'M'N'}) k_{L'M'N'} + k^{*I'J'K'}\\
&+ \frac{1}{2} \ (\frac{1}{2} \ \varphi^{P'Q'[I'} A_{I'P'Q'} \delta_{M'}{}^{J'} \delta_{N'}{}^{K'}
-\varphi^{P'[I'J'}A_{L'M'P'}\delta_{N'}{}^{K']} + \frac{1}{9} \
\varphi^{I'J'K'} A_{L'M'N'}
)k^{*L'M'N'}\\
&-\frac{1}{72\sqrt{3}} \ \epsilon^{I'J'K'P'Q'R'S'T'U'} A_{P'M'N'} A_{L'Q'R'} A_{S'T'U'} k^{*L'M'N'} \ .
\end{split}
\een
In order to obtain these expressions, we had
to use the definitions of $V'$ and of $v'$, perform the Lie-algebra exponential to get
$\V'=\e^{{\rm ad}(v')} {\rm Ad}(V') $, then get $N = \V' \tau'(\V')^{-1}$, from which
$\omega$ is then by definition obtained as $\omega =
\star(K \wedge N^{-1} \D N)$. The Lie-algebra exponential, defined by its infinite power series,
truncates at polynomial order 4 because ${\rm ad}(v')$ is nilpotent of order 5. (In this part of the calculation, we are relying on formulas given in~\cite{mizo1}.) We have also used the geometric condition~\eqref{hsurfo} in the following ways: If we let $g(\xi_8, \xi_{i'}) = w'_{i'}$, then that condition implies $w_{i'}'=0$. In combination with the definitions of the potentials $\chi'_{i'}, \chi'_{i'j' \dots k'}$, there follow
the relations (viewed as relations on $\hat \M=\M/[\mr \times U(1)^8]$):
\ben
\begin{split}\label{threeq}
0&=\D \chi_{i'} + 2 A_{i'j'k'} \D \varphi^{j'k'} - \frac{1}{54} \ \epsilon^{j'k'l'm'n'p'q'r'}
A_{i'j'k'} A_{l'm'n'} \D A_{p'q'r'}\\
0&=\D A_{i'j'}\\
\D A_{8j'k'} &=-\frac{r}{\det f'} \ f_{j'm'}f_{k'n'} \hat \star \D \varphi^{m'n'}
+ \frac{1}{36} \ \epsilon^{i'p'q'r's'l'm'n'} A_{i'p'q'} \hat \star \D A_{r's'l'}
\end{split}
\een
where $\varphi^{i'j'} = \frac{1}{720} \epsilon^{i'j'k'l'm'n'p'q'} \chi_{k'l'm'n'p'q'}'$.
These formulas were used to simplify the expressions
for $k_{I'}{}^{J'}, k_{I'J'K'}, k^{*I'J'K'}$. We may also use~\eqref{scal0} to set
to zero many terms in the expressions for $\omega$. In particular,
the first relation implies, together with the constancy of $\chi_{i'}'$ on $\B$ [cf. \eqref{chi2def}] and of $A_{0i'j'}$, that $U_{i'}=\chi_{i'}' + A_{i'j'k'}\varphi^{j'k'}$ is constant over $\B$.

Next, we evaluate in eqs.~\eqref{j1},~\eqref{j2},~\eqref{j3} the integrals of the 9-forms $\omega_{I'}{}^{J'},
\omega^{*I'J'K'}, \omega_{I'J'K'}$ over the 9-dimensional horizon cross section $\B$, or over the 9-dimensional cross section at infinity. For this, we use the explicit expressions just given. It turns out that the non-vanishing contributions to these surface integrals
consist of surface integrals of the 9-forms $k_{I'}{}^{J'},
k^{*I'J'K'}, k_{I'J'K'}$, multiplied by various constant potentials.
For the surface integrals of the latter 9-forms, one obtains the following expressions.
\ben
\begin{split}
\int_{\B} k_0{}^0 &= 2 \kappa \A \ , \\
\int_{\B} k_i{}^0 &= -J_i \ , \\
\int_{\B} k_i{}^9 &= -J_i \Psi^{jk} \Phi_{jk} \ , \\
\int_{\B} k_0{}^9 &= 4 \ \Psi^{ij}\Phi_{ij} \kappa\A \ ,\\
\int_{\B} k_i{}^j &=0 \ , \\
\int_\infty k_0{}^0 &= \frac{8}{9} \ m  - \Omega^i J_i
\ , \\
\int_\infty k_i{}^0 &=-J_i \ , \\
\int_\infty k_0{}^j &=\delta^{ij} J_i - \frac{8}{9} \ m \Omega^j \ , \\
\int_\infty k_i{}^j &=4\pi \ d^{jm} \tau_{im} + \Omega^j J_i \ , \\
\int_\infty k_i{}^9 &=0 \ ,
\end{split}
\een
where we have not displayed several components that are not needed. It has been used that the constant value of $U_0$ on $\B$ is
$U_0 = - \Psi^{ij} \Phi_{ij}$, by showing that $\chi'_{i'}=O(r^2)$ near $\B$.
This can be seen by integrating the defining relation [compare~\eqref{chi2def}] for $\D \chi'_{i'}=i_{\xi_0} \dots i_{\xi_7} \Q_{\xi_{i'}}$, over a suitable curve $\hat \gamma$ in $\hat \M$
from the horizon to infinity, and by applying the same kind of argument as in the proof of the uniqueness theorem in sec.~\ref{sec:unique}. A similar argument, using the first equation in~\eqref{threeq} then also shows that $U_{i'} = O(r^2)$ near $\B$ for $i'=1,\dots,7$.
We also have
\ben
\begin{split}
\int_\B k_{9jk} &= 2\sqrt{3} \ P_{jk} \ , \\
\int_\B k_{90k} &=0 \ , \\
\int_\B k_{0jk} &= 2\sqrt{3} \ \Phi_{mn}\Psi^{mn} P_{jk} \ , \\
\int_\B k_{ijk} &=0 \ , \\
\int_{\infty} k_{9jk} &= 2\sqrt{3} \ P_{jk} \ , \\
\int_\infty k_{90k} &=0 \ , \\
\int_\infty k_{0jk} &=-2\sqrt{3} \ d_{jm} d_{kn} Q^{mn} \ , \\
\int_\infty k_{ijk} &=0 \ ,
\end{split}
\een
as well as
\ben
\begin{split}
\int_\B k^{*9jk} &= 2\sqrt{3} \ \Phi_{mn} \Psi^{mn} Q^{jk} \ , \\
\int_\B k^{*90k} &= 0 \ , \\
\int_\B k^{*0jk} &= 2\sqrt{3} \ Q^{jk} \ , \\
\int_\B k^{*ijk} &= 0 \ , \\
\int_\infty k^{*9jk} &= -2\sqrt{3} \ d^{jm} d^{kn} P_{mn} \ , \\
\int_\infty k^{*90k} &= 0 \ , \\
\int_\infty k^{*0jk} &= 2 \sqrt{3} \ Q^{jk} \ , \\
\int_\infty k^{*ijk} &=0 \  .
\end{split}
\een
Using these surface integrals in the corresponding surface integrals
for $\omega_{I'}{}^{J'},
\omega^{*I'J'K'}, \omega_{I'J'K'}$ in eqs.~\eqref{j1},~\eqref{j2},~\eqref{j3} yields the
thermodynamic identities quoted in sec.~\ref{sec:massformulas}.

To obtain the above expressions for the surface integrals of the 9-forms
$k_{I'}{}^{J'},
k^{*I'J'K'}, k_{I'J'K'}$, we have used the Komar expressions~\eqref{komar} for
$m, J_i$, which in several cases helps one to read off the interpretation of the
surface integrals at infinity. In several of these expressions, have also used the fact that, at $\B$, we
have
\ben
f_{i'j'} =
\begin{cases}
O(r^2) & \text{if $i'=0$ or $j'=0$,}\\
O(1) & \text{otherwise,}
\end{cases}
\quad
f^{i'j'} =
\begin{cases}
O(r^{-2}) & \text{if $i'=j'=0$,}\\
O(1) & \text{otherwise.}
\end{cases}
\een
These relations follow from the definition of $r$ [cf.~\eqref{rdef}] combined with
the fact that $K$ becomes null on $\B$, and combined with $g(\xi_8, K)=0$ from
eq.~\eqref{hsurfo}. We have also used that $\chi_{i'}' = O(r^2)$ near $\B$.
 We have furthermore used
from the definitions of the electric and magnetic potentials at the horizon~\eqref{phidef}, and~\eqref{psidef}, that
\ben
 \Phi_{jk} = -A_{0jk} \ , \quad
 \Psi^{jk} = -\varphi^{jk} \ .
\een
At infinity, we have also used relations like
\ben
f^{i'j'} = \begin{cases}
-1+O(R^{-1}) & \text{if $i'=j'=0$,}\\
\Omega^i + O(R^{-1}) & \text{if $i'=i, j'=0$,}\\
\delta^{ij}-\Omega^i \Omega^j + O(R^{-1}) & \text{if $i'=i$,$j'=j$,}
\end{cases}
\een
which follow from the asymptotically Kaluza-Klein boundary conditions together with~\eqref{kdef}. We have also used that the electric/magnetic potentials are
of order $O(R^{-1})$ near infinity.

\end{document}